 \documentclass[twocolumn,longauthor,usenames,dvipsnames]{aastex63}
\newbox\grsign \setbox\grsign=\hbox{$>$}
\newdimen\grdimen \grdimen=\ht\grsign
\newbox\laxbox \newbox\gaxbox
\setbox\gaxbox=\hbox{\raise.5ex\hbox{$>$}\llap
     {\lower.5ex\hbox{$\sim$}}}\ht1=\grdimen\dp1=0pt
\setbox\laxbox=\hbox{\raise.5ex\hbox{$<$}\llap
     {\lower.5ex\hbox{$\sim$}}}\ht2=\grdimen\dp2=0pt

\newcommand{\rg}{R_{\rm g}}

\newcommand{\tg}{t_{\rm g}}

\newcommand{\m }{ \mathcal}
\newcommand{\ti }{ \tilde}

\usepackage[T1]{fontenc}
\usepackage{amsmath}
\usepackage{amssymb}
\usepackage{ctable} 
\usepackage{graphicx}
\usepackage{natbib}
\usepackage[titletoc,title]{appendix}
\usepackage{ulem}

\usepackage{mathtools}
\usepackage{lmodern}
\usepackage{csquotes}

\begin{document}
%\tracingall

\title{Relativistic outflows from a GRMHD mean-field disk dynamo}
\shorttitle{A GRMHD disk dynamo}
\shortauthors{Vourellis \& Fendt}

%\correspondingauthor{Christos Vourellis}
\email{vourellis@mpia.de, fendt@mpia.de}

\author[0000-0002-0786-7307]{Christos Vourellis}
\author[0000-0002-3528-7625]{Christian Fendt}

\affiliation{Max Planck Institute for Astronomy, Heidelberg, Germany}

%/////////////////////////////////////////////////////////////////////////
\begin{abstract}
We present simulations of thin accretion disks around black holes investigating a mean-field disk dynamo in our resistive GRMHD code \citep{Vourellis2019} that is able to produce a large scale magnetic flux.
We consider a weak seed field in an initially thin disk, a background (turbulent) magnetic diffusivity and the dynamo action.
A standard quenching mechanism is applied to mitigate the otherwise exponential increase of the magnetic field.
Comparison simulations of an initial Fishbone-Moncrief torus suggest that reconnection may
provide another quenching mechanism.
The dynamo-generated magnetic flux expands from the disk interior into the disk corona, becomes advected by 
disk accretion, and fills the axial region of the domain. 
The dynamo leads to an initially rapid increase in magnetic energy and flux,
while for later evolutionary stages the growth stabilizes.
Accretion towards the black hole depends strongly on the magnetic field structure that develops.
The radial field component supports extraction of angular momentum and thus accretion.
It also sets the conditions for launching a disk wind, initially from inner disk area.
When a strong field has engulfed the disk, strong winds are launched that are predominantly
driven by the pressure gradient of the toroidal field.
For rotating black holes we identify a Poynting flux-dominated jet, driven by the Blandford-Znajek mechanism.
This axial Poynting flux is advected from the disk and therefore accumulates at the expense of 
the flux carried by the disk wind, that is itself regenerated by the disk dynamo.
\end{abstract}

%//////////////////////////////////////////////////////////////////////////////////////


%////////////////////////////////////////////////////////////////////////////////////

\section{Introduction}
One of the fundamental prerequisites for the launching of astrophysical jets is the support of a strong magnetic field. 
A few mechanisms have been proposed that can explain the launching, acceleration, and collimation of these outflows,
all of them further requiring the existence of an accretion disk orbiting a central object (star or black hole).
These processes focus either on jet launching from the surface of an accretion disk via magneto-centrifugal acceleration 
\citep{BP1982,1985PASJ...37..515U, 1986ApJ...301..571P}
or driven by a magnetic pressure gradient \citep{LBell1996},
or from a black hole ergosphere \citep{BZ1977,2005MNRAS.359..801K},
but commonly rely on a substantial magnetic field strength and a favorable magnetic field geometry.

Based on these theoretical ideas, a number of magnetohydrodynamic (MHD) codes were developed, using relativistic gravity,
in order to simulate the rotation of the accretion disk around a black hole and the subsequent jet 
launching \citep{Koide1999, Gammie2003, DeVilliers2003a, Noble2006, DelZanna2007, Porth2017, Ripperda2019ApJS..244...10R}.
Jet launching simulations in the non-relativistic limit have been pioneered by \cite{CK2002} and then
further developed \citep{Zanni2007, MurphyFerreiraZanni2010, Somayeh2012, Stepanovs1, Stepanovs3}.
The usual practice in these works is to apply a strong magnetic field as part of the initial conditions. 
However, the strength and shape of the field may express an unrealistic view of the actual physical case.
Only after the disk-field system has evolved we can say that we have approached a more authentic view of 
the system.
Naturally, there is the downside that the origin and the ab-initio evolution of the jet-launching 
magnetic field is practically unknown.

While for protostellar jets and jets from compact stars the jet launching magnetic field could in principle
be induced by a strong stellar dipolar field (if not advected from the interstellar medium),
this option does not exist for supermassive black holes.
In AGN the jet-driving magnetic field thus must be generated by some kind of {\em accretion disk 
dynamo mechanism}, or be advected from the surrounding medium.
The situation can be different for stellar mass black holes that form by mergers of magnetized 
neutron stars.
Also, the environment of stellar mass black holes resulting form core collapse supernova explosions
could be highly magnetized \citep{2007A&A...474..169C,2018JPhG...45h4001O,2020MNRAS.499.4174M}.

Early studies have derived the structure of black hole magnetospheres in the stationary approach 
\citep{1997A&A...319.1025F, 2000MNRAS.315...89G, 2001A&A...369..308F} 
(but see also \citealt{2017ApJ...836..193P, 2018MNRAS.477.3927M, 2020ApJ...892...37P}),
as well as jet acceleration \citep{2001A&A...369..308F, 2001ApJ...563L.129V, 2003ApJ...592..321T}.
Such steady-state modeling of black hole magnetospheres could be very useful as a basis to study 
particle acceleration and radiation \citep{2018ApJ...868...82T,2019Galax...7...78R}.
While steady state solutions are exact solutions of the physical equations, their stability remains unclear. 
The same holds for the origin of the magnetic field structure considered.

The accretion disk turbulence is strongly believed to be generated by instabilities like the 
magneto-rotational instability (MRI) \citep{BH1991, BH1998}.
At the same time, the MRI is known to amplify the magnetic field, giving rise to a turbulent dynamo effect.
This is typically been investigated using high-resolution, ideal MHD simulations
(see e.g. \citealt{1996ApJ...463..656S,2015ApJ...810...59G})
which are sometimes called {\em direct} dynamo simulations, as the dynamo effect is acting ab initio 
and with no further prescriptions solely based the turbulent motions of the gas as soon as a weak seed field is 
present.

Under certain conditions, these small-scale fluctuations of velocity and magnetic field can be subject to 
a non-linear coupling, which then effectively represents a non-ideal MHD mechanism that amplifies the 
magnetic field on larger scales.
This mechanism is known as {\em mean-field dynamo} \citep{Parker1955, SteenbeckKrause1969a, SteenbeckKrause1969b}
and enters the MHD induction equation as a non-ideal term.
Similarly, the small-scale turbulence effectively results in a turbulent resistivity (or diffusivity)
in the induction equation.
These small-scale turbulent motions may be averaged resulting in a turbulent pattern on larger scales,
typically leading to much stronger coefficients for the  mean-field dynamo effect
and the resistivity if compared to the small-scale
values.

The mean-field dynamo theory is a powerful tool to derive the large scale 
field structure in astrophysical objects.
In the limit of a kinematic dynamo the velocity field is prescribed and remains constant while for
larger magnetic field strength feedback from the field on the flow dynamics is expected and 
the non-linear dynamo theory must be considered.

Overall, it is still not fully clear if such turbulent processes can result in the generation of a well 
ordered large-scale magnetic field that is needed to launch an outflow.
Direct high-resolution simulations typically demonstrate the saturation of the turbulent 
dynamo by showing the magnetic {\em energy} evolution.
However, they will hardly provide an estimate for the production of the large-scale magnetic field. 
What is interesting for jet launching, however, is the generation of a large scale magnetic {\em flux}.
This has been  been achieved by simulations applying the mean-field ansatz in non-relativistic 
studies (see below).
The present paper aims at extending these studies to the GRMHD context.

Concerning the application of the mean-field dynamo theory for accretion disk and jet simulations, 
the literature is rather sparse.
Dynamos have in fact be suggested for generating the magnetic fields needed to launch jets early on.
\citet{Pudritz1981a, Pudritz1981b} have applied the $\alpha\Omega$-dynamo theory for the case of 
thin accretion disks, demonstrating the growth of a magnetic field by means of differential rotation 
of the disk (the $\Omega$-effect), together with the
net effect of turbulent motions (the $\alpha$-effect)
in a vertically stratified turbulent disk.

Simulations of magnetized shear flows by \citet{Brandenburg1995} demonstrated how a dynamo-generated 
magnetic field can amplify the turbulence in the flow, which, in turn, can amplify the magnetic field 
via a dynamo process.
Simulations of outflow launching from a mean-field dynamo active accretion disk
were first presented by \citet{vRekowski2003} and \citet{vRekowski2004}. 
In \citet{ParievColgateFinn2007} a possible origin of the dynamo mechanism in the case of AGN accretion flows
was suggested, triggered by a passing star that can heat up and perturb the magnetic field inside
the accretion disk, altogether resulting in the induction of a toroidal field into a poloidal 
magnetic flux. 

\citet{Stepanovs2} and \citet{FendtGassmann2018} presented simulations of jet launching from accretion 
disk, where the jet driving magnetic field was self-generated by a mean-field disk dynamo.
Different magnetic field structures and thus jet parameters were obtained by exploring 
mean-field dynamos of different strength.
Also, by switching off and on the $\alpha$-dynamo mechanism (externally) a periodic ejection of jets could
be simulated. 
Further, they found that for mean-field dynamos considering a strong $\alpha$, oscillating dynamo modes may occur resulting 
as well in pulsating ejections into the jet (see also \citealt{2018MNRAS.477..127D}).

We now turn to mean-field dynamo models in the relativistic context.
Early on the existence of a gravito-magnetic dynamo effect has been claimed \citep{1996A&A...307..665K},
which could, however, not realized in subsequent studies \citep{1996ApJ...465L.115B, 1996A&A...313.1028K}.
Fully dynamical GRMHD simulations of mean-field dynamos have considered only recently.
Applying GRMHD, \citet{Sadowski2015} parametrized the effect of a mean-field dynamo in combination with radiative transfer.
However, their approach was in ideal MHD regime neglecting magnetic diffusivity.
Also a simplified dynamo model was used, where the dynamo action appears as small numerical correction for the
evolution of poloidal magnetic field that drives it towards a saturated state.
They applied their code to investigate the evolution of thick disks at different accretion rates. 

The first fully covariant implementation of a dynamo closure in a general relativistic context was 
presented by \citet{BdZ2013} following the 3+1 formalism. 
A mean-field dynamo was implemented in the \texttt{ECHO} code \citep{DelZanna2007, BucciantiniDelZanna2011},
that was later applied to simulations of {\em kinematic} dynamo action in tori around black holes
demonstrating the growth of the toroidal and poloidal field components \citep{Bugli2014}.
Recently, \citet{Tomei2020MNRAS.491.2346T} have extended these studies by simulating fully dynamical cases including dynamo quenching. 

In our present paper, we aim for a broader description of the mean-field dynamo in GRMHD simulations.
In comparison to the previously mentioned work we consider in particular the following points.

(i) Compared  to \citet{Bugli2014} in our paper we do consider the feedback of the magnetic field on 
the dynamics of the system. 
A fully dynamical MHD study brings a major advance over a kinematic study, and is particularly  essential concerning disk accretion and jet launching.

(ii) Compared to \citet{Tomei2020MNRAS.491.2346T} who consider the  dynamical feedback by the field 
and also discuss the distribution of the poloidal field energy, we will investigate in detail the geometry 
of the poloidal field that is generated including the overall structure of the magnetic field
lines. This type of analysis was not performed by \citet{Bugli2014}.

(iii) Compared to \citet{Tomei2020MNRAS.491.2346T} we also show the evolution of hydrodynamical quantities.
This is in particular essential for a discussion of the disk accretion and the jet launching.

We apply our {\em resistive} GRMHD code \texttt{rHARM3D} \citep{QQ1,QQ2,Vourellis2019} and
run a series of fully dynamical simulations with an accretion disk mean-field dynamo.
In particular, we aim to demonstrate the growth of the poloidal magnetic field, thus the
generation of a large scale magnetic flux that is needed to launch winds or jets from the accretion disk.

Our paper is structured as follows.
In Section~\ref{sec:tb} we review the basic GRMHD equations, discuss the mean-field dynamo including its implementation as well as its quenching.
In Section~\ref{sec:torustest} we apply the dynamo effect for the setup of a relativistic torus, which allows us to compare
and test our results with the existing literature.
In Section~\ref{sec:tddynamoinit} we detail the initial conditions for our setup of a thin disk mean-field dynamo.
In Section~\ref{sec:tddynamosims} we present simulation results, discussing the evolution of the magnetic field and the
the accretion disk, the mass fluxes, and the launching of disk winds and relativistic jets.
In Section~\ref{sec:sum} we summarize our work.

%=========================================================================================================
\section{Theoretical background}
\label{sec:tb}
In this section, we briefly review the basic equations of resistive GRMHD as a basis for our calculations. 
For the metric we adopt the signature ($-,+,+,+$) \citep{MTW1973} and apply geometrized units, $G=c=1$. 
Thus, length scales are expressed in units of the gravitational radius $\rg = GM/c^2$,
while time is measured in units of the light travel time $\tg = GM/c^3$.
Vector quantities are written with bold letters while the vector and tensor components are indicated with their respective indices, with Greek letters running for 0,1,2,3 ($t,r,\theta,\phi$) and Latin letters running for 1,2,3 ($r,\theta,\phi$). 

The usual $3+1$ decomposition for GRMHD is used in order to separate the time component from the spatial components
(3-dimensional manifolds).
The space-time is described by the metric $g_{\mu \nu}$ in Kerr-Schild coordinates with $g \equiv det(g_{\mu \nu})$.
A zero angular momentum observer frame (ZAMO) exists in the spacelike manifolds, moving only in time with the
velocity $n_{\mu} = (-\alpha,0,0,0)$, where $\alpha = 1/ \sqrt{-g^{tt}}$ is the lapse function. 
The gravitational shift is $\beta^i = \alpha^2 g^{ti}$.

In our simulations we apply our resistive code rHARM3D \citep{Vourellis2019}, now extended by including 
a turbulent mean-field dynamo, following the work of \citet{BdZ2013}.

%--------------------------------------------------------
\subsection{Basic GRMHD equations}
The Maxwell equations in covariant form 
\begin{equation}
     \nabla_{\nu} F^{*\mu \nu} = 0, \quad\quad
     \nabla_{\nu} F^{\mu \nu} = J^{\mu},
\label{eq:Maxwell}
\end{equation}
along with the conservation of the stress-energy tensor
\begin{eqnarray}
   \nabla _{\nu} T^{\mu \nu} = 0,
\label{eq:motion}
\end{eqnarray}
describe the motion of a magnetized fluid in a general relativistic environment. 
$J^{\mu}$ is the 4-current that satisfies the electric charge conservation 
$\nabla_{\mu} J^{\mu} = 0$, $T^{\mu \nu}$ is the stress-energy tensor and $\nabla_{\mu}$ is the covariant derivative.
The electromagnetic field is described by the Faraday and Maxwell tensors,
\begin{equation}
\begin{split}
F^{\mu \nu} & =  u^{\mu}e^{\nu} - e^{\mu}u^{\nu} + \epsilon^{\mu \nu \alpha \beta} u_{\alpha} b_{\beta},
 \\
F^{*\mu \nu} & =  - u^{\mu}b^{\nu} + b^{\mu}u^{\nu} + \epsilon^{\mu \nu \alpha \beta} u_{\alpha} e_{\beta},
\end{split}
\label{eq:Faraday}
\end{equation}
respectively,
where $u^{\mu}$ is the 4-velocity and $e^{\mu}$ and $b^{\mu}$ express the electric and magnetic field in the fluid 
rest frame. The Levi-Civita symbol $\epsilon^{\mu \nu \kappa \lambda}$ is defined as
\begin{equation}
\epsilon_{\alpha \beta \gamma \delta} =             \sqrt{-g}[\alpha \beta \gamma \delta], \quad
\epsilon^{\alpha \beta \gamma \delta} = - \frac{1}{\sqrt{-g}}[\alpha \beta \gamma \delta],
\label{eq:LevCiv}
\end{equation}
where $[\alpha \beta \gamma \delta]$ are the conventional permutation symbols.
The magnetic and electric field as measured by the zero angular momentum observer (ZAMO) are defined as
$ \m B^{\mu} =  - n_{\nu} F^{*\mu \nu} $ and $ \m E^{\mu} = n_{\nu} F^{\mu \nu}$, with $ \m B^i = \alpha F^{*it} $, $ \m E^i = - \alpha F^{it}$, $ \m B^0= \m E^0=0$.
The stress-energy tensor $T^{\mu \nu}$ 
can be split into a fluid and an electromagnetic component.
The fluid component is
\begin{equation}
    T^{\mu \nu}_{\textrm fluid} = (\rho + u + p)u^{\mu}u^{\nu} + pg^{\mu \nu}, 
\label{eq:Tfluid}
\end{equation}
where $\rho$ is the mass density, $u$ is the internal energy density and $p$ is the thermal pressure. 
Pressure and internal energy are connected through the equation of state for an ideal gas with
\begin{equation}
u = \frac{p}{\Gamma -1},
\label{eq:eos}
\end{equation}
where $\Gamma$ is the polytropic exponent. 
The electromagnetic component of the stress-energy tensor is
\begin{equation}
\begin{split}
    T^{\mu \nu}_{\textrm EM} = & (b^2 + e^2) \left(u^{\mu}u^{\nu} + \frac{g^{\mu \nu}}{2} \right) - b^{\mu}b^{\nu} - e^{\mu}e^{\nu} \\ & - u_{\alpha}e_{\beta}b_{\gamma} \left( u^{\mu}\epsilon^{\nu \alpha \beta \gamma } + u^{\nu}\epsilon^{\mu \alpha \beta \gamma } \right)
\label{eq:Tem}
\end{split}
\end{equation}
Equations \eqref{eq:Maxwell}, \eqref{eq:motion} along with \eqref{eq:eos} are being solved by our code as a system of hyperbolic differential equations. For a more detailed analysis of the equations and their numerical implementation and solution we refer the reader to \citet{BdZ2013}, \citet{QQ1} and \citet{Vourellis2019}.

%-----------------------------------------------------------------------------
\subsection{The mean-field dynamo}
In ideal MHD, the electric field vanishes in the co-moving frame which is expressed as
\begin{equation}
    \pmb{E} +  \pmb{v} \times \pmb{B} = 0, 
\end{equation}
where $\pmb{v}, \pmb{E}, \pmb{B}$ are the velocity and the electric and magnetic field.
In the more general case of non-ideal MHD, Ohm's law is given by 
\begin{equation}
    \sigma \pmb{E'} = \sigma \left(\pmb{E} + \pmb{v} \times  \pmb{B} \right) =  \pmb{J},
\end{equation}
where $\pmb{E'}$ is the co-moving electric field, $\pmb{J}$ is the electric current and $\sigma$ is the electric conductivity.

In the mean-field ansatz
the turbulent fluctuations of velocity and magnetic field lead to an averaged, mean electromotive force,
\begin{equation}
    \overline{ \delta\pmb{v} \times \delta \pmb{B} } = \alpha_{\rm D} \pmb{B}  - \beta_{\rm D} \pmb{J},
    \label{eq:EMF}
\end{equation}
that subsequently enters the induction equation.
Here, $\alpha_{\rm D}$ describes the term that is responsible for the generation of the magnetic field
from the turbulent motions, the so-called $\alpha$-effect,
while the $\beta_{\rm d}$ provides an additional resistive term, that enhances the dissipation of the magnetic 
field by an enhanced turbulent diffusivity in addition to the ohmic resistivity
\citep{Parker1955, SteenbeckKrause1969a, SteenbeckKrause1969b}.

We can thus write the induction equation as
\begin{equation}
    \frac{\partial \pmb{B}}{\partial t} = \nabla \times (\pmb{v} \times \pmb{B}) + \nabla \times \alpha \pmb{B} - \nabla \times \eta\pmb{J},
    \label{eq:alpha-induct}
\end{equation}
where we now have defined $\alpha \equiv \alpha_{\rm D}$ and the magnetic diffusivity $\eta$ considering
both the turbulent and ohmic components.
Renaming the coefficients also emphasizes the fact that the mean-field dynamo theory starts from the {\em ideal} MHD when 
considering the correlated turbulent fluctuations, but after performing the averaging of the electromotive force term, an extra term
for the {\em non-ideal} treatment of MHD appears.

The first term on the r.h.s. of the induction equation represents a kinematic term that may induce a poloidal magnetic field from a toroidal
motion, the so-called $\Omega$-effect.
Combined with the action of the $\alpha$-effect, this may provide a dynamo cycle that induces a poloidal field from a toroidal field
($\alpha$-effect) that then induces a toroidal field from the poloidal component ($\Omega$-effect), and so forth - the so-called
$\alpha\Omega$-dynamo.
Note that also the $\alpha$-effect may induce a toroidal field from a poloidal component.
This is in particular important in stellar dynamos with small shear in the toroidal motion, thus forming an $\alpha^2$ dynamo.
However, for strong shear, for example as in accretion disks following a Keplerian rotation or in relativistic tori,
the $\Omega$-effect will dominate the $\alpha\Omega$-dynamo concerning the induction of the toroidal field component.

In a more general definition of the mean-field electromotive force, 
the $\alpha_{\rm D}$ and the $\beta_{\rm D}$ will represent {\em tensors} \citep{Moffatt1978}.
However, in most applications (see e.g. \citealt{Stepanovs2,FendtGassmann2018,2018MNRAS.477..127D}), 
including the present paper, they are treated as scalar functions.
An exception is \citet{Mattia2020} who perform simulations of dynamo action and jet launching considering
an anisotropic dynamo alpha.
Note that if $\alpha$ and $\eta$ are spatially constant, they can be moved in front of the curl operator in 
Equation~\eqref{eq:alpha-induct}.

We want to briefly connect to the {\em direct} dynamo simulations that were mentioned in the introduction.
Those simulations, applying high resolution, actually may resolve the turbulence pattern in the MHD flow 
and thus perform the averaging equation~\ref{eq:EMF}, thus providing an $\alpha_{\rm D}$ and  $\beta_{\rm D}$,
that can be than used for simulations in the mean-field approach.
As an example, we cite here \citet{Gressel2010}, who applied accretion disk shearing box simulations in order
to derive profiles for the magnetic diffusivity and the dynamo coefficients.
As complementary to the mean-field approach, the direct simulations serve as {\em subgrid modeling},
that can deliver the mean-field quantities from a subgrid scale.

%------------------------------------------------------------------
\subsection{The GRMHD mean-field dynamo equations}
The implementation of the mean-field dynamo mechanism builds up on our previous works in which we have 
integrated resistivity in the form of turbulent diffusivity in the original HARM code, establishing 
our resistive GRMHD code \texttt{rHARM3D} \citep{QQ1, QQ2, Vourellis2019}.

For our GRMHD treatment, we follow the closure relation introduced by \citet{BdZ2013}.
Here, the mean-field $\alpha$-dynamo parameter is replaced by $\xi = -\alpha$, just to avoid confusion 
with the gravitational lapse.
Thus, in covariant form, in the fluid frame Ohm's law is written as
\begin{equation}
    e^{\mu} = \eta j^{\mu} \, +\; \xi  b^{\mu},
\label{eq:theo:grmhd:Ohm1}
\end{equation}
where $j^{\mu}$ is the 4-vector of the electric current density.
Now, the electric field can no longer be calculated by the cross product of fluid velocity and magnetic field 
and new equations need to be formulated (see next Section).

By setting $\xi=0$ we get the resistive version of Ohm's law $\left(e^{\mu} = \eta \,j^{\mu}\right)$.
By setting $\eta=0$ we get back to the ideal MHD case $e^{\mu}=0$.

In order to derive the equations for the evolution of the magnetic and electric field we start by taking 
the temporal and spatial projections in Equation~\eqref{eq:Maxwell} separately.
This provides the divergence condition $\partial_j \left(\gamma^{1/2} B^j\right) = 0$
and an equation for the time evolution of the magnetic field,
\begin{equation}
\begin{aligned}
        & \gamma^{-1/2}\partial_t \left(\gamma^{1/2} \m B^i \right) +
        \epsilon^{ijk} \partial_j \left( \m E_k \alpha  +  \epsilon_{knm} \m B^m \beta^n \right)
        = 0,
        \label{eq:Bevol}
\end{aligned}
\end{equation}
along with Gauss' law for the electric field $\partial_j \left(\gamma^{1/2} \m E^j\right) = \gamma^{1/2} \,q$ and the time evolution of the electric field,
\begin{equation}
\begin{aligned}
        \gamma^{-1/2}\partial_t \left(\gamma^{1/2} \m E^i \right) -
        \epsilon^{ijk} \partial_j &\left( \alpha \m B_k  -  \epsilon_{knm} \m E^m \beta^n \right)= \\
        &= \left(q\beta^i - \alpha J^i\right).
        \label{eq:theo:grmhd:Amprere2} 
\end{aligned}
\end{equation}
where $\gamma$ is the determinant of the spatial metric $\gamma_{\mu \nu} = g_{\mu \nu} + n_{\mu} n_{\nu}$.
From  Ohm's law, Equation~\eqref{eq:theo:grmhd:Ohm1}, we can get
\begin{equation}
    F^{\mu\nu} u_{\nu} = \eta I^{\mu} + \eta \left(I^{\nu}u_{\nu} \right)u^{\mu} + \xi F^{*\mu\nu}u_{\nu},
\end{equation}
where $I^{\mu}$ is 4-current as is seen by the ZAMO.
\citet{BlackmanField1993} derived this equation only for the case of a resistive fluid, 
however, the addition of a mean-field dynamo term is straightforward.
The source term $I^{\nu}u_{\nu} =  - q_0$ is the electric charge density as measured in the fluid frame \citep{Komissarov2007}.

By taking the temporal decomposition of Equation~\ref{eq:theo:grmhd:Ohm1} we get an expression for the electric charge in the form
\begin{equation}
    \Gamma \m E^{i}v_{i} = \eta (q - \Gamma q_0)  + \xi\Gamma \m B^{i}v_{i},
    \label{eq:tmpDec}
\end{equation}
and from the spatial projection of Equation~\ref{eq:theo:grmhd:Ohm1} we get an expression for the electric current
\begin{equation}
\begin{aligned}
      \Gamma\left[ \m E^i  + \epsilon^{ijk} v_j \m B_k - (\m E^kv_k)v^i \right] = \eta \left(J^i - q v^i \right) + \\
      \xi\Gamma \left[ \m B^i - (\m B^kv_k) v^i - \epsilon^{ijk} v_j \m E_k  \right],
\end{aligned}
\label{eq:sptDec}
\end{equation}
which we can now use to replace the source terms in Equation~\eqref{eq:theo:grmhd:Amprere2}, resulting in the evolution equation for the electric field

\begin{equation}
\begin{aligned}
    \gamma^{-1/2} \partial_t &\left(\gamma^{1/2} \m E^i \right) -
    \epsilon^{ijk} \partial_j \left( \alpha \m B_k  -  \epsilon_{knm} \m E^m \beta^n \right) = \\
    &= -\alpha\Gamma\left[ \m E^i  + \epsilon^{ijk} v_j \m B_k - (\m E^kv_k)v^i \right]/\eta \\
    & + \alpha\xi\Gamma \left[ \m B^i  -  \epsilon^{ijk} v_j \m E_k   -  (\m B^kv_k) v^i \right] /\eta\\
    & - q( \alpha v^i - \beta^i).
    \label{eq:dyn:evolE}
\end{aligned}
\end{equation}

Discretizing the time evolution of the electric field leads after some \enquote{minor} algebraic calculations to
\footnote{We note a correction here, namely the differently placed parentheses for the term on the r.h.s of 
the $1^{\rm st}$ line, and the different sign for the term on the $4^{\rm th}$ line in comparison to \citet{BdZ2013}.}
\begin{equation}
\begin{aligned}
  \m E^{i} &\left[\ti\eta + \Gamma + \xi^2 \frac{ (\Gamma^2 - 1) }{\ti\eta + \Gamma} \right] = \\
    & - \epsilon^{ijk} \ti v_j  \m B_k 
      + \ti\eta \left[Q^i + \frac{\ti v^i (Q^k\ti v_k)}{\Gamma\ti\eta + 1} \right] \\
    & + \xi\left[ \Gamma \m B^i - \frac{\ti\eta \ti v^i}{\Gamma\ti\eta + 1}(\m B^k\ti v_k) \right]  \\
    & - \xi \left\{
    \frac{\ti\eta\epsilon^{ijk}\ti v_j Q_k +  (\Gamma^2 - 1) \m B^i  -  (\ti v^j \m B_j)\ti v^i }{\Gamma + \ti\eta} \right\}\\
    & - \xi^2 \frac{\Gamma\epsilon^{ijk}\ti v_j \m B_k}{ \Gamma + \ti\eta } \\
    & + \xi^2 \ti v^i \left\{ \frac{\Gamma\ti\eta Q^k\ti v_k  
      +  \xi \m B^k\ti v_k }{(\Gamma\ti\eta + 1)(\Gamma + \ti\eta)}  \right\},
\end{aligned}
\label{eq:num:dynamoE}
\end{equation}
where
\begin{subequations}
\begin{align}
        Q^i &= \m E^i_0 + \Delta t \left[-\left(\alpha v^i - \beta^i\right)q + 
        \epsilon^{ijk} \partial_j \left(\alpha \m B_k - \epsilon_{klm} \beta^l \m E^m \right) \right],
        \label{eq:num:nonstiff} \nonumber\\
        q &= \gamma^{-1/2} \partial_k \left(\gamma^{1/2} \m E^k \right), \nonumber \\
        \tilde{\eta} &= \frac{\eta}{\alpha \Delta t}, \nonumber \\
        \ti v^i &= \Gamma v^i. \nonumber
\end{align}
\end{subequations}
with $\m E^i_0$ representing the electric field as calculated in the previous time step.

Unfortunately, Equation~\eqref{eq:dyn:evolE} appears to be {\it stiff.} 
For low values of diffusivity, the terms on the r.h.s. evolve in different times than the timestep used for the time 
discretization of of the electric field, resulting in unstable solutions.
The problem can be fixed by isolating the {\it stiff} terms and use an implicit method to calculate them.
The non-stiff part of the electric field, that is the one which does not include diffusive terms,
is calculated separately, and is denoted as $Q^i$.
For the {\it stiff} part we use an iterative method where for every time step we calculate the electric field until its values have converged to the desired accuracy.
A detailed analysis of the numerical implementation is presented in \citet{QQ1}.

We clearly state that we have tested the implementation of resistivity considering analytical, time-dependent solutions 
of the induction equation \citep{QQ1, Vourellis2019} and find perfect agreement.
Now, since the additional implementation of a mean-field dynamo is achieved in exactly the same way as for the resistivity,
so our testing of resistivity supports the implementation of the dynamo term as well.

%------------------------------------------------------------------------------------------------------------------------------------
\subsection{Dynamo quenching}
The exponential increase of the magnetic field strength is a direct result of the mean-field dynamo process.
However, a strong magnetic field will suppress the MRI which is considered as one of the sources of the disk turbulence.
Therefore, with the repression of turbulence, also the dynamo will be suppressed, resulting in a saturation level 
of the magnetic field generated.
This has been demonstrated by direct simulations of the MRI \citep{1996ApJ...463..656S, Gressel2010, 2015ApJ...810...59G}
and has been approved analytically \citep{Ruedigeretal1995}.
The supression of MRI has recently be considered also in GRMHD \citep{2018MNRAS.475..108B}.

In our approach, in absence of a more detailed turbulence model, the dynamo quenching must be 
enforced by interactively lowering the value of the dynamo parameter, $\xi$, following certain 
physical criteria, motivated by models of turbulence.

The usual procedure (that we refer to as {"}standard quenching{"}) prescribes a kind of equipartition field strength 
beyond which
the mean-field dynamo parameter is reduced, until the dynamo becomes effectively inactive.
For example, \citet{Bardou2001} have implemented a back-reaction on the $\alpha$-dynamo parameter from the magnetic field 
by making this alpha $\alpha$ depend on the values of $\bm B$
(see also \citealt{vRekowski2003, vRekowski2004} ).
A different approach was undertaken by \citet{FendtGassmann2018} who applied a disk diffusivity model where the strength 
dynamo was quenched by increasing the magnetic diffusivity. 
This resulted an a smooth and long-term evolution for their jet launching simulations (up to several 100000 disk rotations).

However, the rate in which the magnetic diffusivity and the mean-field dynamo affect the magnetic field evolution
can be very different.

A leading parameter of mean-field dynamo action is the dynamo number.
For the case of a thin accretion disk of scale height $H$, we apply the dynamo parameter $\xi$ and magnetic diffusivity $\eta$ for
defining the {\em total dynamo number} $\cal D_{\xi}$, following previous works in a non-relativistic environment utilizing a thin accretion disk \citep{Bardou2001, vRekowski2003, Stepanovs2},
\begin{equation}
    {\cal D}_{\xi} = \left|{\cal R}_{\xi}{\cal R}_{\Omega} \right| = \left| \frac{\xi H}{\eta} \frac{S H^2}{\eta} \right|
    \label{eq:dynamonumber}
\end{equation}
where the two fractions represent the Reynold's numbers ${\cal R}$ from the dynamo action turbulence 
(mean-field $\alpha$-dynamo)
and from differential rotation ($\Omega$-dynamo), respectively.
The function $S = r \;d\Omega / dr$ measures the shear due to differential rotation and is calculated for each cell separately.

Depending on the values of the dynamo number, $\xi$ will exponentially increase the magnetic field, 
while $\eta$ would damp with a rate close to linear.
Thus, if the dynamo number is too large, the change in diffusivity as suggested by \citet{FendtGassmann2018} might not be 
sufficient and a standard quenching model must also be employed.
Furthermore, by increasing the diffusivity, the numerical time step drastically decreases, increasing the computational costs
of the simulation.

In our models, we follow the standard quenching as e.g. applied by \citet{Bardou2001}.
We define the plasma-$\beta$ as the ratio between the gas and the magnetic pressure and magnetization $\mu$ as its inverse

\begin{equation}
    \beta = \frac{1}{\mu} = \frac{p_{\text{gas}}}{p_{\text{magn}}}.
\end{equation}
We implement a quenching prescription that calculates the magnetization $\mu$ of the fluid and when it
becomes too large, the actual dynamo parameter $\xi_{\textrm q}$ is reduced,
\begin{equation}
    \xi_{\textrm q} = \xi_0 \frac{1}{1 + \mu_{\textrm D}/\mu_{\textrm eq}},
    \label{eq:ch3:quenching}
\end{equation}
where $\xi_{\textrm q}$ is the quenched dynamo parameter, $\xi_0$ its initial value, $\mu_D$ the disk magnetization {\em averaged vertically} over the disk
and $\mu_{\textrm eq}$ is a equipartition magnetization, expressing the value of the disk magnetization 
in which the mean-field dynamo is quenched by 50\% (i.e. $\xi_{q} = \xi_0/2$).

Note that initially we encountered an issue with the MPI parallelization of the code.
In order to take average values for the magnetization over a certain region of cells to calculate the 
actual quenching, these cells must actually belong in the same processor.
Alternatively, we would have to find a way to communicate the average values of certain disk areas between the processors.
The second option would require heavy work on the parallelization of HARM, so we follow the first option.
The downside of our choice is a restriction to a small number of cores for our simulations, just due to the need 
of covering a large area of the disk by a single core.

We compromise by running the simulations with only a few number of cores in the $\theta$ direction so that the 
calculation of the quenching is run by the same core in the angular direction.
For our $256 \times 256$ grid we use 64 cores with 4 cells each in the radial direction and 5 cores with 51 
cells each in the polar direction resulting in a total of 320 cores.

%=========================================================================================================
\section{The dynamo effect in a relativistic torus}
\label{sec:torustest}
In this section we investigate the mean-field dynamo effect in a relativistic torus.
This setup is essentially different from that of a thin accretion disk in the sense that the differential rotation
of a torus is {\em stronger} compared to a disk, thus the $\Omega$-effect of the mean-field dynamo plays a  stronger
role. 
This becomes clear if we compare the dynamo numbers for the cases of the disk and the torus.
In a torus with a constant specific angular momentum the angular velocity scales as $\propto r^{-2}$ while 
for a Keplerian rotating disk the scaling follows $\propto r^{-3/2}$. 
We assume a scale height $H_{\rm D} = 0.1 R$ for a thin disk 
as typical length scale for the disk dynamo action.
Similarly, as typical length scale for the torus dynamo we take the $e-$folding length scale $H_{\rm T}$ of the density 
distribution.
Considering the centre of the torus ($r=15$), the first contour is at $z\approx 4.2$, so $H_{\rm T}/R = 0.285$.
We thus find from Equation~\eqref{eq:dynamonumber}  the Reynold's numbers from the differential rotation
\begin{eqnarray}
   {\cal R}_{\rm T} = R\frac{\partial \Omega_{\rm T}}{\partial R} \frac{H_{\rm T}^2}{\eta} = \frac{0.16}{\eta}, \\
   {\cal R}_{\rm D} = R\frac{\partial \Omega_{\rm D}}{\partial R} \frac{H_{\rm D}^2}{\eta} = \frac{0.015}{\eta} \sqrt{R}.
\end{eqnarray}
Note that the dynamo number of the torus can be higher by an order of magnitude in the inner area,
for larger radii, however, the disk dynamo number increases.

We can then write
\begin{eqnarray}
  {\cal R}_T /{\cal R}_D = 
    \frac{0.16}{0.015\sqrt{R}} \approx 10 R^{-1/2}.
\end{eqnarray}
This relation changes of course with time as the disks evolve.
Compared to these systems with strong shear, stellar dynamos are better described
by a $\alpha^2$ mean-field dynamo, and are found to be 
less stable and less symmetric \citep{1999A&A...346..922K}.

\begin{figure}
     \centering
    \includegraphics[width=0.45\columnwidth]{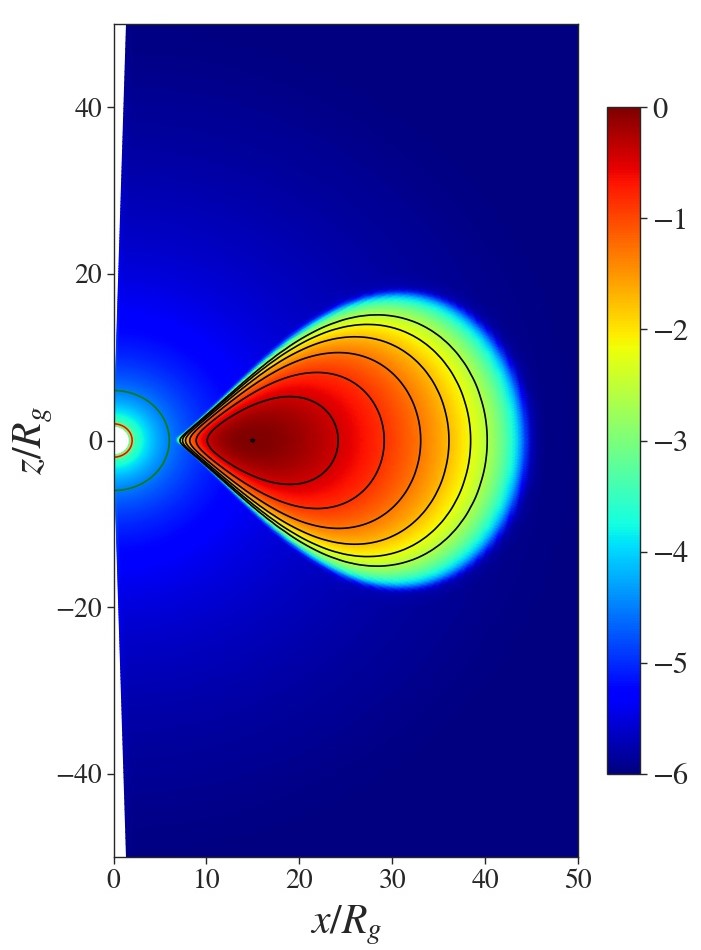}
    \includegraphics[width=0.45\columnwidth]{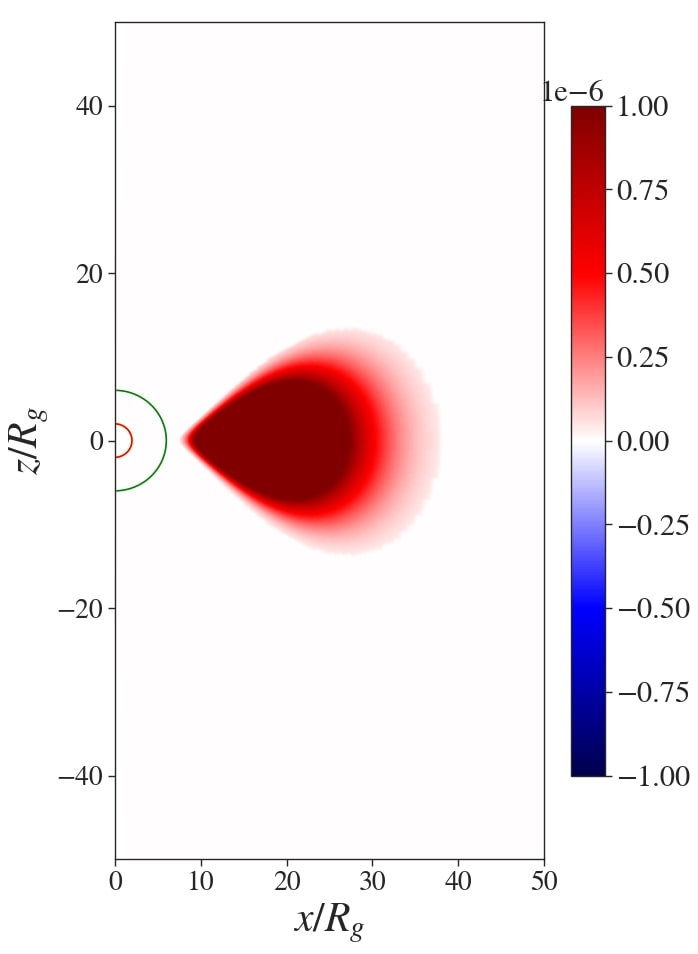} \\
    \includegraphics[width=0.45\columnwidth]{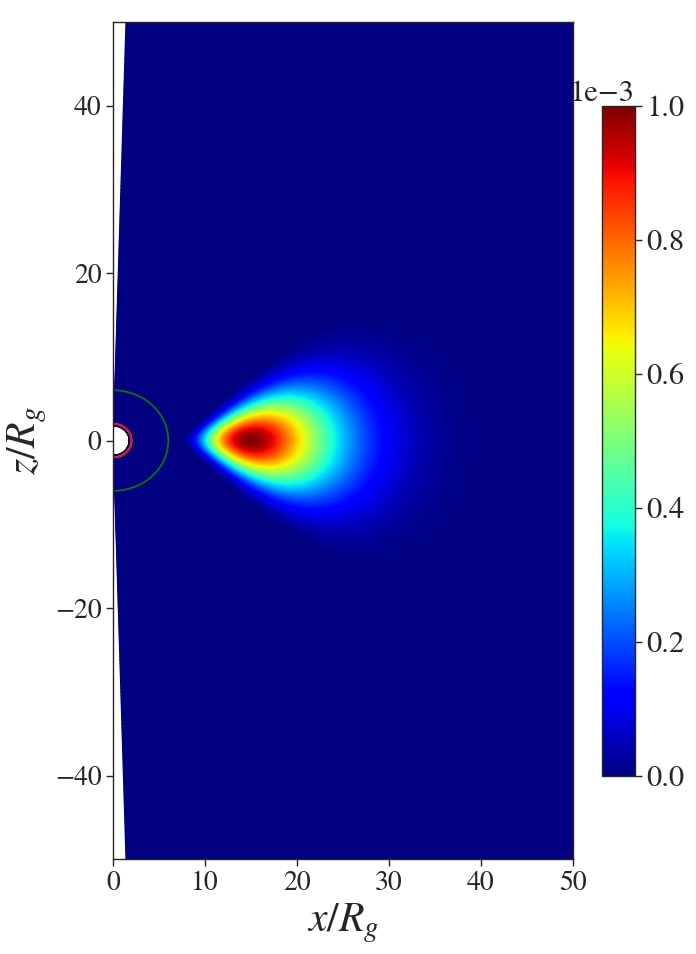}
    \includegraphics[width=0.45\columnwidth]{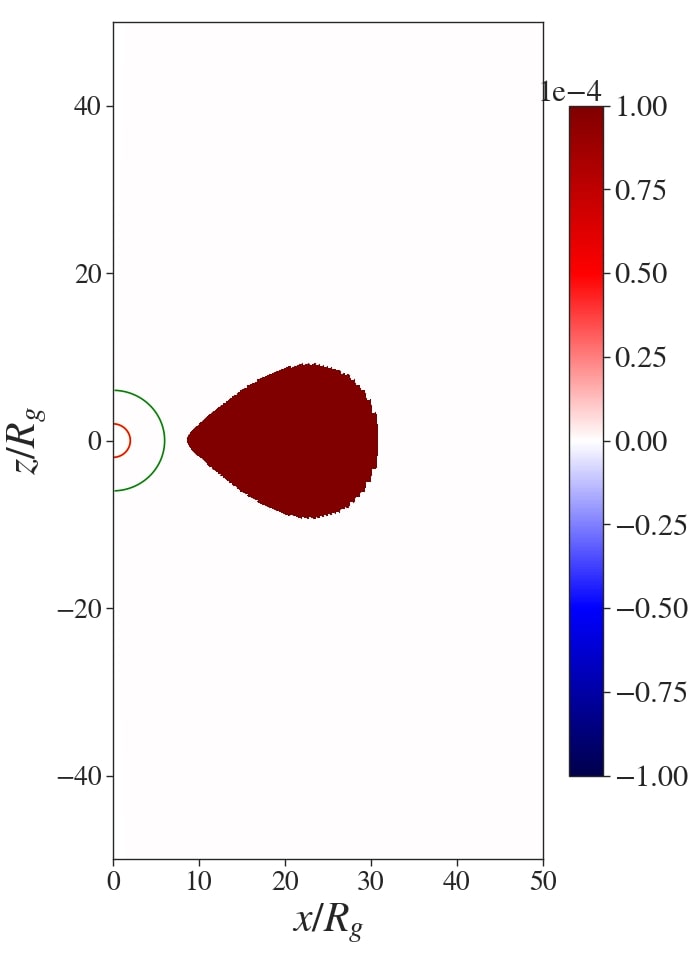}
    \caption{\small Initial conditions for our reference simulation. 
      Shown is the initial distribution of the density, including contours separated by one  $e-$folding scale height
      (logscale; upper left), 
      the toroidal magnetic field $B_{\phi}$ (linear scale; upper right), 
      the diffusivity $\eta$ (linear scale; lower left), 
      and the dynamo parameter $\xi$ (linear scale; lower right). }
     \label{fig:ch3:initial_torus}   
 \end{figure}

 \begin{figure}
     \centering
     \includegraphics[width=0.45\columnwidth]{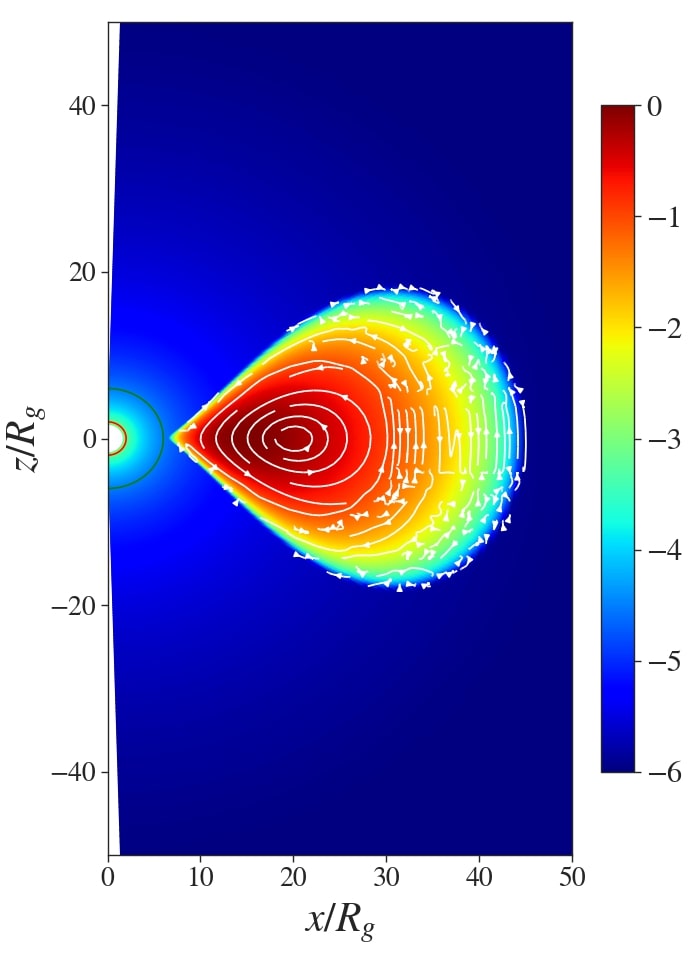}
     \includegraphics[width=0.45\columnwidth]{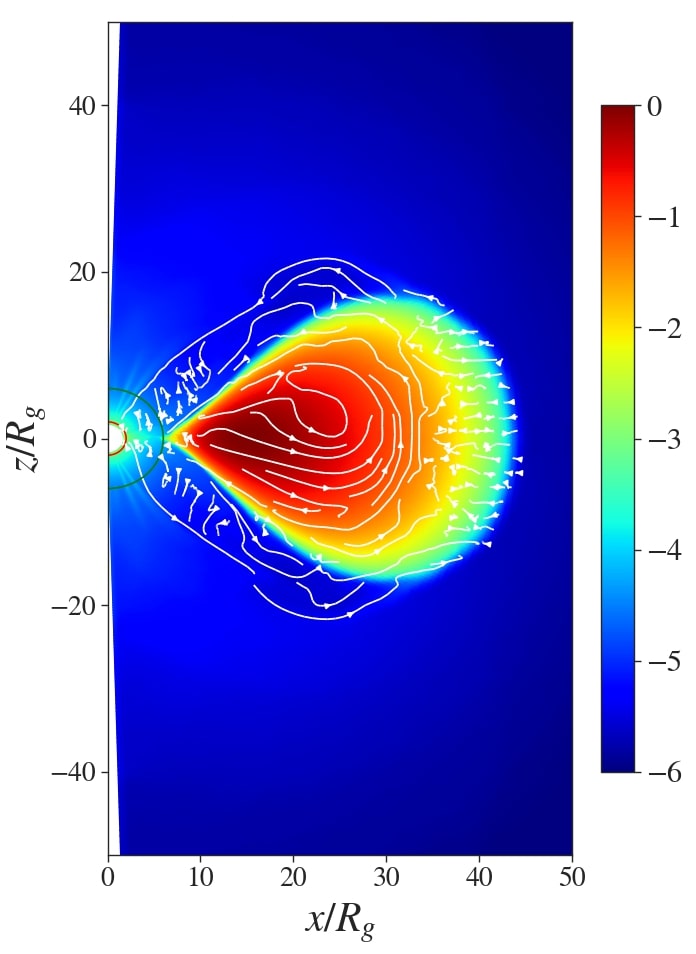} \\
     \includegraphics[width=0.45\columnwidth]{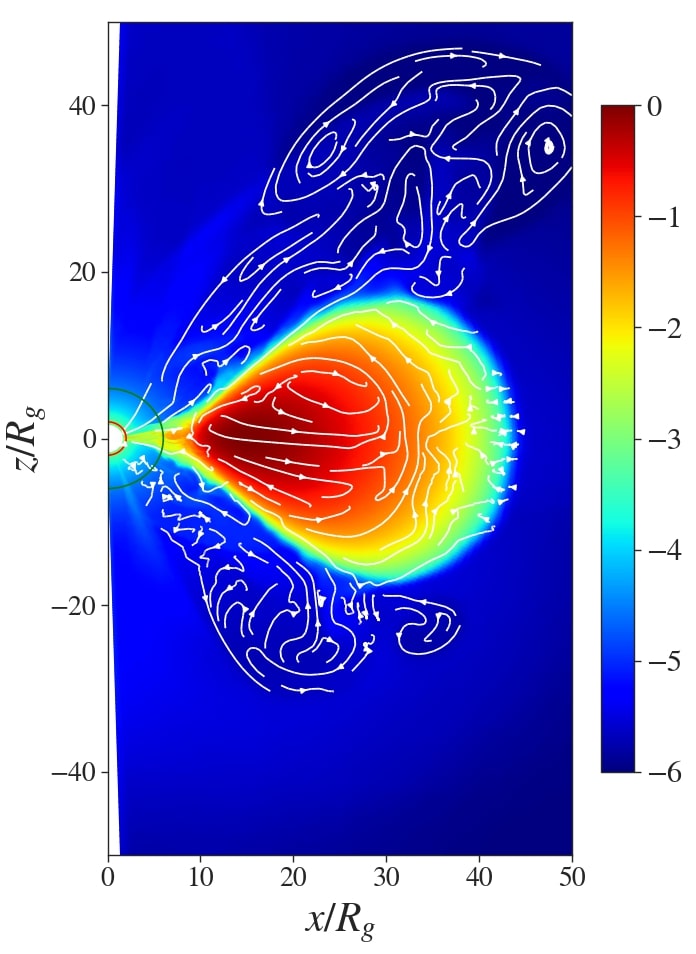}
     \includegraphics[width=0.45\columnwidth]{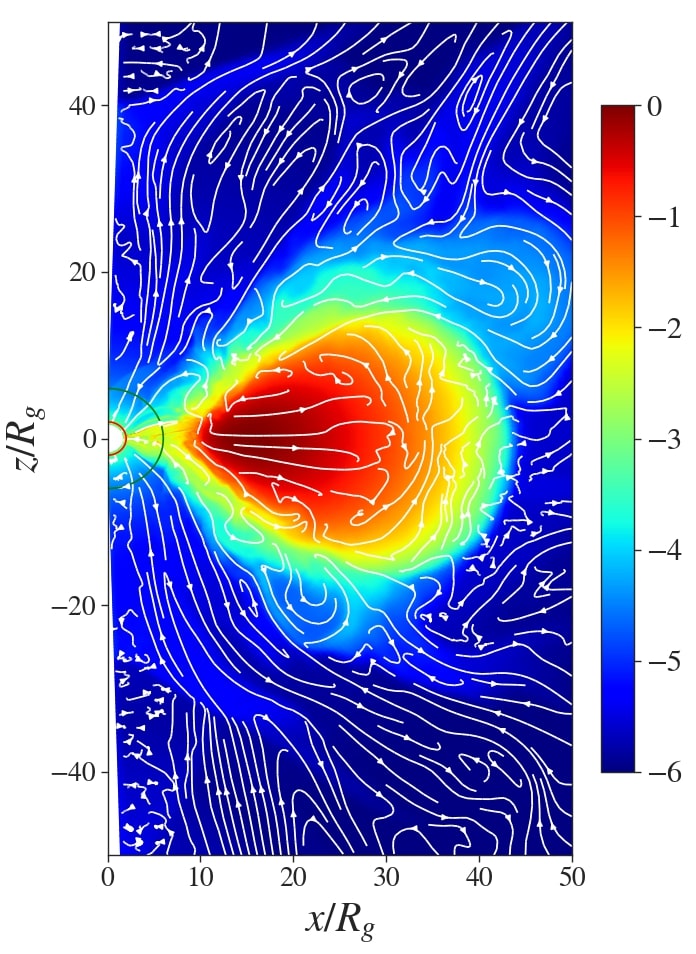}
      \caption{\small Generation of a dipolar poloidal field. The color gradient shows the density distribution in log scale and the white lines show the poloidal magnetic field. 
      The initial condition starts with a toroidal field with high plasma-$\beta \sim 10^6$. The poloidal field appears immediately in the area where dynamo exists ($t=1$, upper left) and it evolves through $t=1000, 2000$ and $3500$. This corresponds in approximately 10 rotations of the point of maximum density at $R_{\rho_{\text{max}}}=15$. }
     \label{fig:ch3:rholines1}
 \end{figure}

Our simulation setup for this exemplary torus simulation {\em simT} is as follows.
We start the simulation with a density and internal energy distribution following 
 the Equation of State for an ideal gas, which reads as
\begin{equation}
    \rho = \left( \frac{(h-1)(\gamma -1)}{K\gamma} \right)^{1/(\gamma-1)},
\end{equation}
where $h(r,\theta)$ is the specific enthalpy as it is calculated in \citet{FishboneMoncrief1976}.
The torus is rotating around a Schwarzschild black hole ($a=0$) with an inner edge
at $r_{\text{in}}=6$ and its point of maximum density at $r_{\rho_{\text{max}}}=15$.
The internal energy is defined by the polytropic equation of state $u = K\rho^{\gamma}/{\gamma -1}$,
with $K=10^{-3}$ and $\gamma = 4/3$.

An initial toroidal magnetic seed field is defined inside the torus, following the density 
distribution ($B_{\phi} \propto \rho$) of the standard Fishbone-Moncrief torus applied in HARM \citep{Gammie2003}.
This guarantees that the magnetic field is confined inside the torus.
The field direction is positive in both hemispheres and it is normalized considering a plasma-$\beta_0=10^6$.

We prescribe a distribution for the magnetic diffusivity that follows the density distribution of
the torus with a maximum value of $\eta_0 = 10^{-3}$.
The diffusivity basically vanishes in the surrounding corona.
Also the dynamo action is limited to the area inside the torus.
It is constrained to an area smaller than distribution of the initial magnetic field and the
diffusivity in order to avoid field generation in the corona of the torus.
The dynamo parameter is constant in both hemispheres at a level of $\xi_0 = 10^{-3}$.

The parameters of our torus simulation {\em simT} are summarized in Table~\ref{tab:torus}.
The initial conditions for this simulation are shown in Figure~\ref{fig:ch3:initial_torus}.

We note that this setup is somewhat different from the one chosen by \citet{Tomei2020MNRAS.491.2346T}, 
as their initial torus is that of a so-called {\em Polish donut} (see e.g.~\citealt{2013LRR....16....1A}) 
and their seed field geometry is that of poloidal loops inside the torus.
With our approach we prefer to connect our simulations 
to the literature of HARM simulations which typically apply the Fishbone-Moncrief torus as initial condition, and 
While our initial setup may look similar to \citet{Tomei2020MNRAS.491.2346T} our results cannot directly compared.

\begin{table}[t]
    \caption{Summary of parameters of the torus simulation run. 
    Listed is the ID of the run; 
    the Kerr parameter $a$ applied; 
    the maximum magnetic diffusivity $\eta_0$; 
    the initial plasma-$\beta$
    and the dynamo parameter $\xi_0$.
    }
    
    \centering
    \begin{tabular}{ccccc}
    \hline
    \noalign{\smallskip}
        run   &  $a$  & $\eta_0$ & $\beta_0$ & $\xi_0$\\
    \noalign{\smallskip}     
    \hline 
    \noalign{\smallskip}
        {\em simT} &  0  & $10^{-3}$ &  $10^6$ & $10^{-3}$  \\
    \noalign{\smallskip}
    \hline
    \end{tabular}
    \label{tab:torus}
\end{table}

%---------------------------------------------------------------------------------------
\subsection{Field induction by a spatially constant dynamo}
We now describe the field evolution of a torus induced by a mean-field dynamo in more detail.

When the simulations starts, the poloidal magnetic field appears immediately inside the torus in
the area where $B_{\phi}$ and dynamo coexist.
As the simulation continues, the field starts increasing in value due to the dynamo while advecting towards the black 
hole following partially the accretion of the torus increasing the magnetic field in the inner disk part and towards the black hole.
Simultaneously, diffusivity is trying to dampen the magnetic field. 
In the end, the ${\cal R}_{\xi}$ determines where the field is amplified or dampened.
Since $\xi$ is constant, the profile of diffusivity determines the value of $R_{\xi}$. 
As an example of their values, in the center of the torus the numbers are $R_{\xi} \sim 15$ while its maximum values reaches $R_{\xi} \sim 150$ in the boundaries of the dynamo distribution.

 \begin{figure*}
     \centering
     \includegraphics[width=0.55\columnwidth]{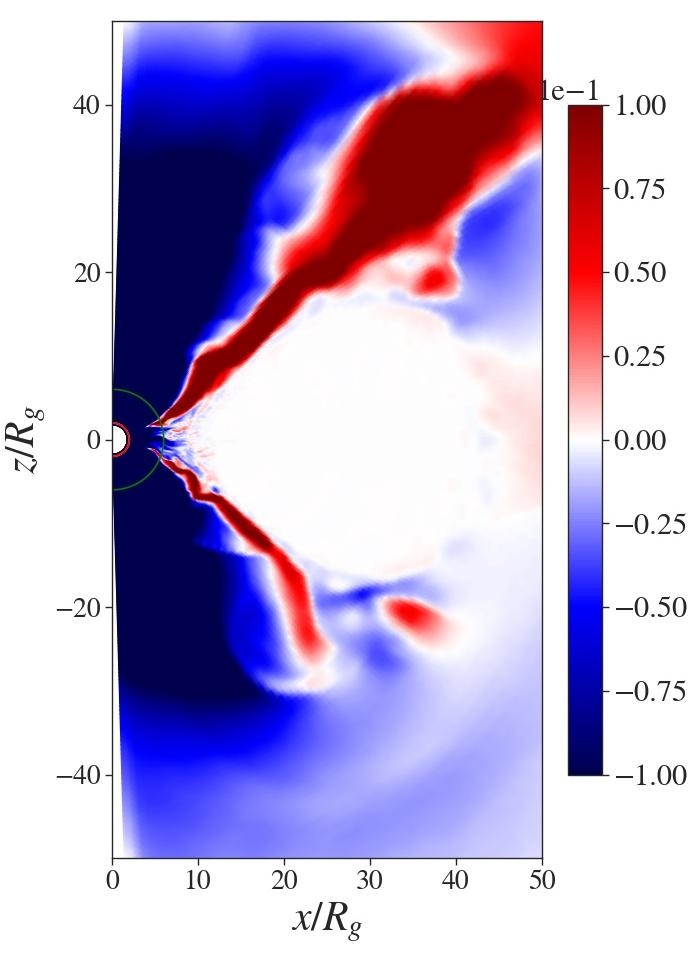}
     \includegraphics[width=0.55\columnwidth]{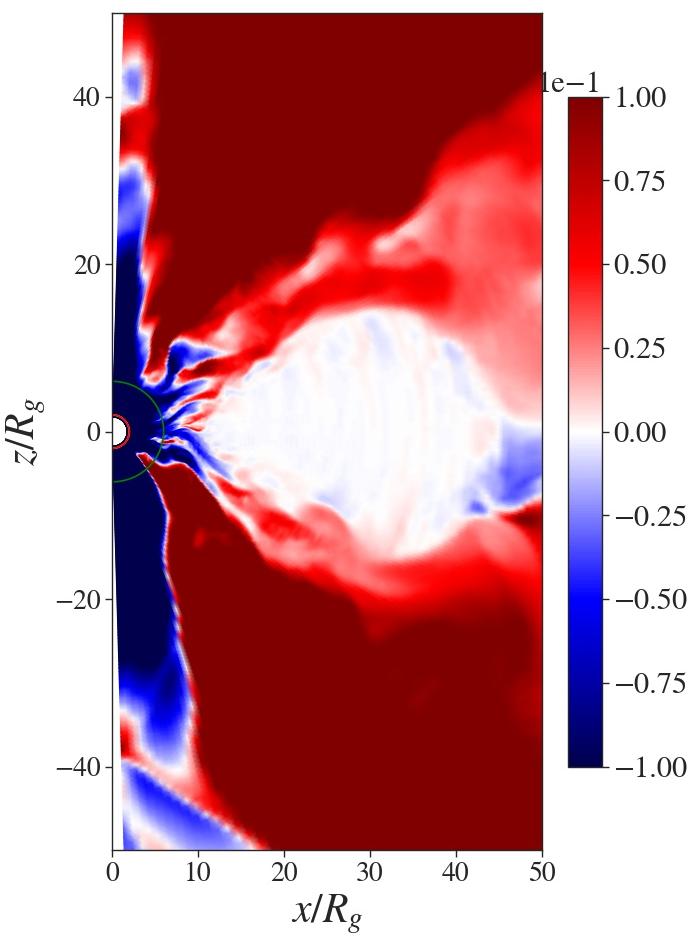}
     \includegraphics[width=0.55\columnwidth]{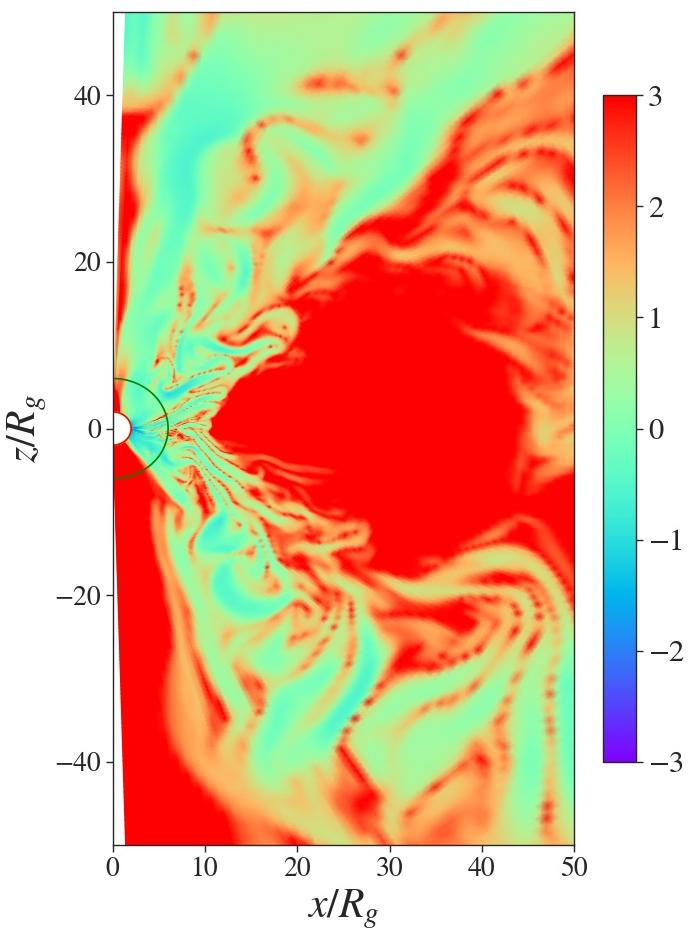}
      \caption{Distribution of the radial velocity at $t=2000$ (left) and $t=3800$ (middle) 
       as well as plasma-$\beta$ at $t=3800$ (right)
       in the torus simulation.
       The color bars are chosen to highlight the 4 different velocity areas, such as the torus (white), 
       the funnel flow (dark red), the torus wind (light red), and the infall (dark blue), as well as the areas
        of high (green) and weak (red) magnetization.}
     \label{fig:ch3:beta_vp}
 \end{figure*}

 \begin{figure}
     \centering
     \includegraphics[width=0.95\columnwidth]{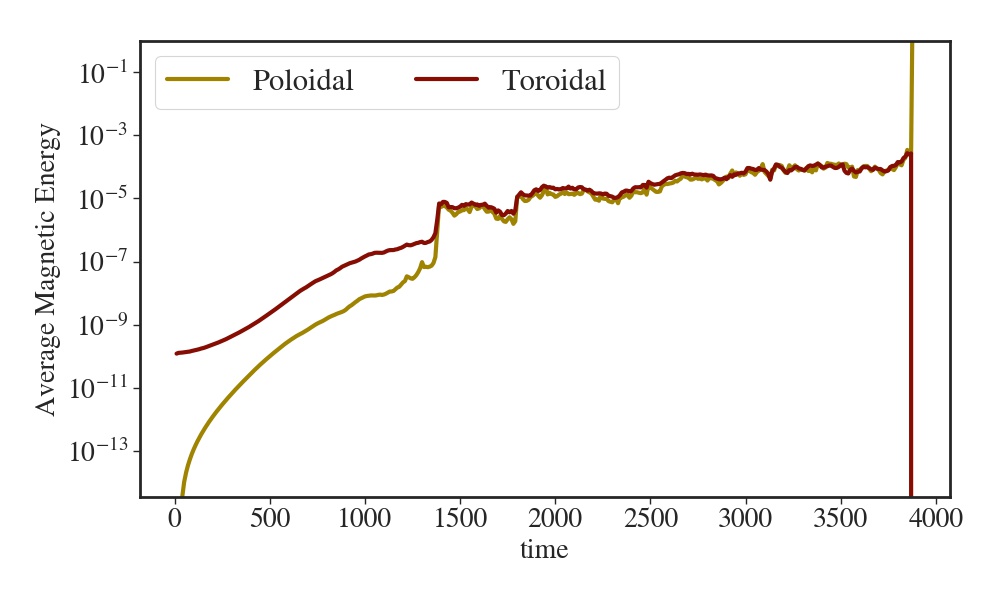}
      \caption{Time evolution of the average magnetic energy across our simulation grid for the poloidal and 
      toroidal field component. }
     \label{fig:ch3:bp_bphi}
 \end{figure}

In Figure~\ref{fig:ch3:rholines1} we see the snapshots of the evolution of the poloidal magnetic field.
As  mentioned before, in the beginning, the field lines are restricted in the area where $\xi \neq 0$.
However, they are eventually dragged along with the material that is accreted towards the black hole.
At time $t=2000$, as the dynamo has been working for approximately 6 rotations of the torus center, we see
a low velocity outflow being launched from the inner part of the torus.
The magnetic field lines follow the low density fluid showing the first indications of the development of a jet.
The launching point of the outflow is barely inside of the innermost stable circular orbit (ISCO) and can be attributed to the presence of the strong toroidal magnetic field we see in the close atmosphere of the inner part of the torus.
This strong toroidal field, which has been amplified by the $\Omega$ effect of the dynamo, can increase the magnetic pressure and push material and poloidal field outwards \citep{LBell1996}.

The poloidal field continues to grow up to $t \sim 3800$, but then the field close to the axis becomes too strong, 
resulting the code to fail to converge.
At that time, both the toroidal and the poloidal components of the field have spread into the greatest part of the grid while the 
plasma-$\beta$ has reached values between $0.01 - 0.1$ inside the ISCO and around 1-10 in the outflow.
Still, the torus is hydrodynamically dominated (see Figure~\ref{fig:ch3:beta_vp}).

The material accelerated in the outflow consists mainly of the floor values used by the code as background environment.
We observe outflow velocities larger than $0.1\,c$ (see Figure~\ref{fig:ch3:beta_vp}).
The acceleration is supported by magnetic pressure forces. 
This holds for the high speed floor material, but for the torus material that moves with a radial speed $\simeq 0.01$ (the light-red area in Figure~\ref{fig:ch3:beta_vp}, left) we expect that the driving is in addition also supported by gas pressure.
Note also that along the axis where the toroidal field vanishes we observe infall towards the black hole.

In Figure~\ref{fig:ch3:bp_bphi} we show the evolution of the average magnetic energy for both the poloidal and the toroidal field.
In both channels the energy increases over orders of magnitude until a saturation level appears around $t=1500$.
There is no full saturation as beyond $t=1500$ the field energy still increases by a factor of 100-1000, however
the strong increase of the dynamo action flattens substantially.
The flattening is interesting insofar as we do not apply dynamo quenching in this simulation.
So the dynamo seems to self-quench.
This has also been seen in the literature (see below).
We hypothesize that this is due to re-connection of the tangled magnetic field with time, as Figures~\ref{fig:ch3:rholines1} and~\ref{fig:ch3:beta_vp}
show clear evidence for an increasingly turbulent state of the magnetic field.
Due to the application of resistivity, this field may physically reconnect, lowering the magnetic energy and heating the plasma. 

We think that the process works in two ways.
First, the dynamo-generated magnetic field can transported out of the dynamo-active area, however an area where
there is still a non-negligible resistivity (basically the light red area in 
Fig.~\ref{fig:ch3:initial_torus} (upper right). 
Since the field is tangled (see Figures~\ref{fig:ch3:rholines1}) it will reconnect.
However, the same process may happen also inside the torus. 
The resistivity is largest inside the torus (dark red areas in Fig.~\ref{fig:ch3:initial_torus}, right),
thus, reconnection will be very efficient here. 
Considering this and also that the dynamo-generated field is strongest here (the $\xi$ is large),
reconnection will work efficiently in lowering the net magnetic field energy.

Interestingly, this points towards a new possible channel for dynamo quenching by reconnection.
Considering magnetic diffusivity, we may mostly think of diffusing away magnetic flux and by this lowering the efficiency of the
dynamo.
However, also reconnection is a result of magnetic diffusivity, that does naturally lowers the field energy.

%----------------------------------------------------------------------------------------------------
\subsection{The dynamo-generated torus magnetic field}
The initial condition of the toroidal component of the magnetic field follows the density profile while keeping a positive sign in 
both hemispheres.
A toroidal field is also produced by the $\Omega$ effect by differential rotation of the induced poloidal field, increasing 
the initial toroidal field.

The induced toroidal field changes sign in the equatorial plane, keeping its positive values 
in the upper hemisphere (where the radial field is negative).
This results in a gradual change in the toroidal field across the torus, 
when the initial condition is superimposed by the newly generated field.

At time $t=1000$, the sign has changed already within the borders of the area where dynamo activity has 
prescribed.
The change of sign of the toroidal field then continues to gradually affect also the lower hemisphere.
At time $t=2000$ this transformation has almost finished, resulting in a toroidal field distribution that
changes sign across the equatorial plane.

The dynamo-generated poloidal field is dominated by the radial component.
Note that the direction of the radial field component also affects the direction of the 
poloidal field.
Here we have the problem that the dynamo-generated field which is initially prescribed as 
a perfect dipole (with negative values in the upper hemisphere and positive values in 
the lower hemisphere), eventually turns into a form of stripes that show alternating 
positive and negative directions, which eventually result in the formation of closed 
magnetic loops.

This effect of inducing a layered structure of the radial magnetic field components
naturally also affects the direction of the toroidal field, where we see a similar behavior 
in later evolutionary stages.
This effect becomes apparent from $t=500-800$.
At this time, the initial dipolar field geometry begins to change its topology, 
as we find a positive $B_r$ in the upper hemisphere, that was initially set to a negative
radial field structure.

Similarly, this is seen in all components of the magnetic (and also in the electric field).
From $t=1000$ an additional layer appears in the $B_r$ distribution, now with a negative 
value that is alternating with the previously induced positive $B_r$ in the upper hemisphere.

We observe this process for the whole duration of the simulation run and 
eventually find a completely layered magnetic field structure of alternating field direction.
We note that this layered structure shows up preferentially close to the equatorial plane
of the torus.  
Here, we also find high dynamo numbers of the turbulent dynamo exactly within
previously mentioned layer.
In later times this region of layered magnetic field direction spreads out over the whole torus
and can eventually be found in the outflows as well.
This dynamical process was also reported in \citet{Bugli2014} and \citet{Tomei2020MNRAS.491.2346T} 
despite a different prescription for the symmetry of the dynamo parameter.

The time evolution and the spatial structure of the magnetic field evolving strongly reminds
on dynamo waves that are generated during the kinematic phase of the mean field dynamo process.
Dynamo waves are typically discussed as oscillatory solutions of the {\em kinematic} dynamo problem.
The existence of such dynamo waves have been investigated early by \citet{Parker1955}
and many subsequent studies (see e.g. \citealt{1984GApFD..30..305W, 2010rdla.book...15H}).

Here, we solve for the dynamical dynamo problem, i.e. essentially considering the back reaction
of the magnetic field generated on the gas.
Nevertheless, in the early stages of the simulations, we are still in the linear regime
of the dynamo action, and the field is not yet strong enough (high plasma-$\beta$) to act on 
the gas. 
The situation is thus comparable to the kinematic regime and the excitation of dynamo waves may indeed be expected.
More recent studies have detected also dynamo waves in cases of high magnetic Reynolds numbers
\citep{2014ApJ...789...70C,2017MNRAS.464L.119N}.
This is essential to know, as also in our simulations the magnetic Reynolds numbers 
${\rm R_{\text m}} \equiv (v L) / \eta \simeq 1000 $ can be high, assuming as a typical length scale the torus
scale height $L \simeq 10$,
a typical velocity $v=0.1$, and a maximum diffusivity of $\eta = 0.001$ (all in code units).
We note here that much of the work cited here has been done for spherical stellar dynamos.
However, the case of of the relativistic torus is probably not too far from this, as compared
to our studies of thin disks later on in this paper, the shear in the torus is relatively weak. 

\begin{figure}
     \centering
     \includegraphics[width=0.6\columnwidth]{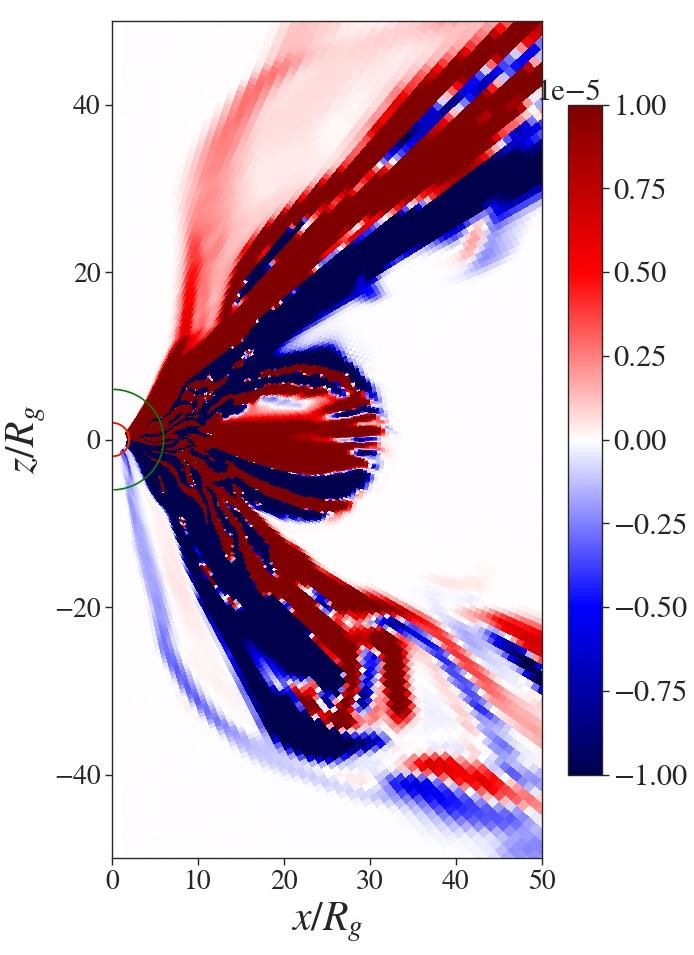}
      \caption{The radial magnetic field structure in simulation {\em simT} at time $t=2800$. }
     \label{fig:br-waves}
\end{figure}

A characteristic feature of dynamo waves is that the typical time scale of the fluctuations
is similar to the diffusion time scale (see e.g. \citealt{2005AN....326..693G} for spherical $\alpha^2$-dynamos).
For the simulations presented in this section, this is indeed the case.
With a maximum diffusivity $\bar{\eta} \simeq 10^{-3}$ in the torus\footnote{Of course this value quite 
changes over the area of the torus} and the typical length scale in the torus $\Delta R \simeq 1$,
we calculate a global diffusion time scale of $\tau_{\rm diff} = (\Delta R )^2 / \bar{\eta} \simeq 10^{3}$
in code units. 
This is indeed similar to the time scale we measure for the global variations of the dynamo-induced field
structure.
Although the values considered are some rough estimates and vary drastically over the torus, we may see this
correlation as a hint for dynamo waves in our simulations.

Even at late stages of our simulation, the plasma-beta in the torus is of the order of 100-1000,
thus the dynamo still acts in the kinematic regime.
Nevertheless, the generated magnetic field has expanded and fills the space between the rotational
axis and the torus.
Here, the plasma-beta is lower $\beta \simeq 0.1$, and the evolution is dominated by the magnetic field.

Overall, we see indication that dynamo waves are excited during the initial stages of the magnetic field 
evolution on our torus model simulation.
This is in particular demonstrated in Figure~\ref{fig:br-waves} that shows the generation
of a layered structure of an anti-aligned field (here shown by the by the radial field component.
This figure show striking similarities to Figure~3 in \citet{Tomei2020MNRAS.491.2346T}, who, however, plot the
magnitude of the field only, not its direction.

\subsection{Comparison to literature studies}
At this point it is interesting to compare to the existing litrerature, in particular to the simulations of \citet{BdZ2013} 
on which the mean-field dynamo closure of our code is based and the follow-up papers \citep{Bugli2014, Tomei2020MNRAS.491.2346T} . 
As mentioned above, while the geometry of the latter two approaches look similar, the actual initial conditions
are quite different.
Our initial torus structure follows the Fishbone-Moncrief solution \citep{FishboneMoncrief1976} that is applied
in previous HARM simulations, while their torus follows a different prescription.
Overall, our initial torus extends to $r\simeq 45$, while \citet{Bugli2014} consider a quite smaller structure with $r<10$.
The {\em Polish donut} torus solution applied in \citet{Tomei2020MNRAS.491.2346T} appears to be much larger, however
of the full grid of $r<100$ only the innermost radii $r<20$ are shown.

Also the initial field structure and the diffusivity distributions are different.

Given the different initial setup a quantitative comparison to our results is not possible.
However, we observe similar structures evolving from our dynamo.
Similar to \citet{Tomei2020MNRAS.491.2346T} we observe the generation of bipolar magnetic {"}arms{"} around the inner edge of
the torus.
Also, authors claim to detect dynamo waves being dragged towards the black hole by accretion.

We show the 2D magnetic field evolution at a somewhat longer time step
(in codes units $t=3500$ vs. $1608$ in \citealt{Tomei2020MNRAS.491.2346T}).
Therefore we are able to follow the expansion of the generated magnetic flux away from the torus and towards the 
rotational axis.
At $t=3500$ we indeed see a strongly magnetized axial region, that may allow for Blandford-Znajek jets in case of
a rotating black hole.

We note however that the dynamical evolution of the torus and also of the dynamo action is quite faster
in their smaller sized torus setup.
The 6 periods of torus orbits they show in their 2D images (and the 12 orbits they actually simulated) 
are calculated considering a radius of the torus center of $r_c =12$, while the center of our torus
is further out $r_c =15$, resulting in about a factor 2 longer rotational periods for our torus center
and thus a similarly longer evolution time.
The advection time of magnetic flux towards the axis is not affected by that, however.

Interestingly, \citet{Tomei2020MNRAS.491.2346T} find a (second) saturation phase of field amplification even without
quenching their mean-field dynamo, probably due to the accretion dynamics together with an equipartition state 
of the fluid system

In our simulations we also find a transition to a saturation state, more accurately, a strong break in the growth rate of both field
components
(see Figure~\ref{fig:ch3:bp_bphi} and our corresponding discussion above).
\citet{Tomei2020MNRAS.491.2346T} argue that the change is happening at the time when local equipartition is reached.
We may confirm this as close to the inner edges of the torus we also find equipartition at this stage 
(see Figure~\ref{fig:ch3:beta_vp} upper right for a latter evolutionary stage).
However, we hypothesize that the change in slope for the increase of magnetic energy can also be due by reconnection of
the tangled magnetic field that is produced by the dynamo action (see discussion above).

We do not find the second saturation phase claimed by \citet{Tomei2020MNRAS.491.2346T}.
The time scale of our simulation would be sufficiently long, however, since our simulation setup is different, in particular
the dynamics of initial torus, a proper comparison cannot be made here.
We note again, that measured in orbital periods of the torus, the evolutionary time of their torus is 
somewhat longer, and we might therefore have not yet reached the second saturation phase (if this phase would be present
at all in our setup).

The poloidal field structure derived by \citet{Tomei2020MNRAS.491.2346T} is visualized by showing the
magnetic energy distribution.
Here, we also show poloidal field lines that can better trace the small-scale
structure of the field, as well as the field direction.

To see the very {\em structure} of the vector field, meaning the geometry of the magnetic field lines 
and the distribution of the magnetic flux, is essential for understanding the jet launching process\footnote{Note that while the poloidal magnetic flux $\int \vec{B}_{\textrm p} d\vec{A}$ may vanish on average, 
the magnetic energy $\propto B_{\textrm p}^2$ can still be substantial. }.
For example, for considering magneto-centrifugally driven disk winds, the field inclination is
important \citep{BP1982}.

So far, for the test simulations provided in the literature of GR-MHD dynamos 
\citep{BdZ2013, Bugli2014, Tomei2020MNRAS.491.2346T},
field lines for the resulting magnetic field structure are only given for the case of a neutron star dynamo 
(see Section 5.2.4 of \citealt{BdZ2013}).
Thus we cannot compare our field structure in detail with the existing literature.
This would have been interesting as the essential effect of the dynamo action is in particular to
generate poloidal magnetic flux along the disk.

%--------------------------------------------------------------------------------------------
\section{Initial setup for a thin disk dynamo}
\label{sec:tddynamoinit}
Here we describe the initial conditions for our dynamo simulations considering a thin accretion disk.
The simulation grid extends from just inside the event horizon to a outer radius of $r = 80$.
The radial and polar dimensions are based on the original grid of \texttt{HARM}
The numerical grid size used is $256 \times 256$.

\begin{figure*}
\centering
    \includegraphics[width=0.65\columnwidth]{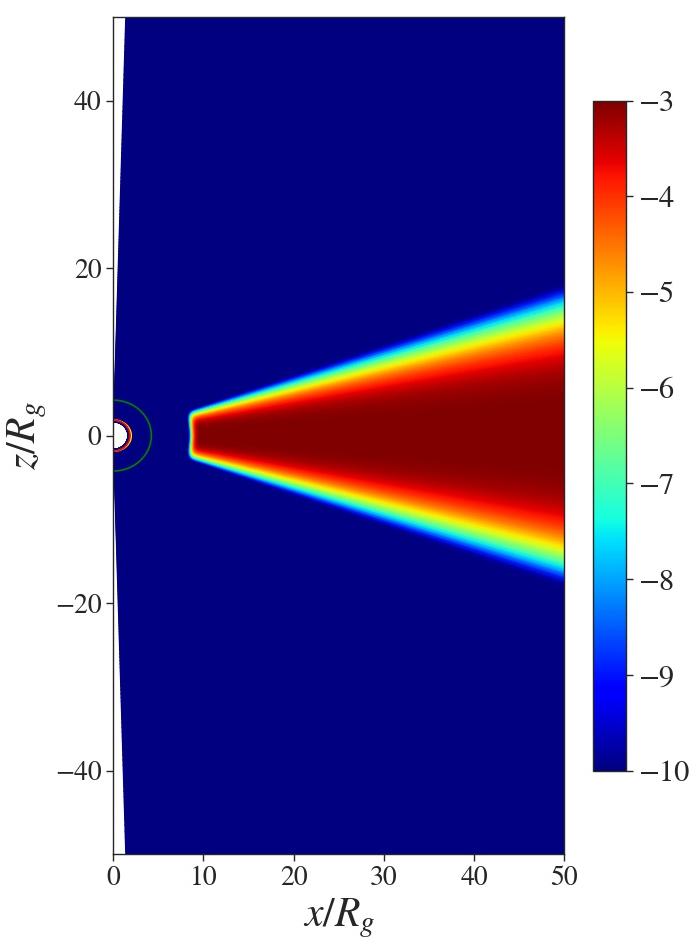}
    \includegraphics[width=0.65\columnwidth]{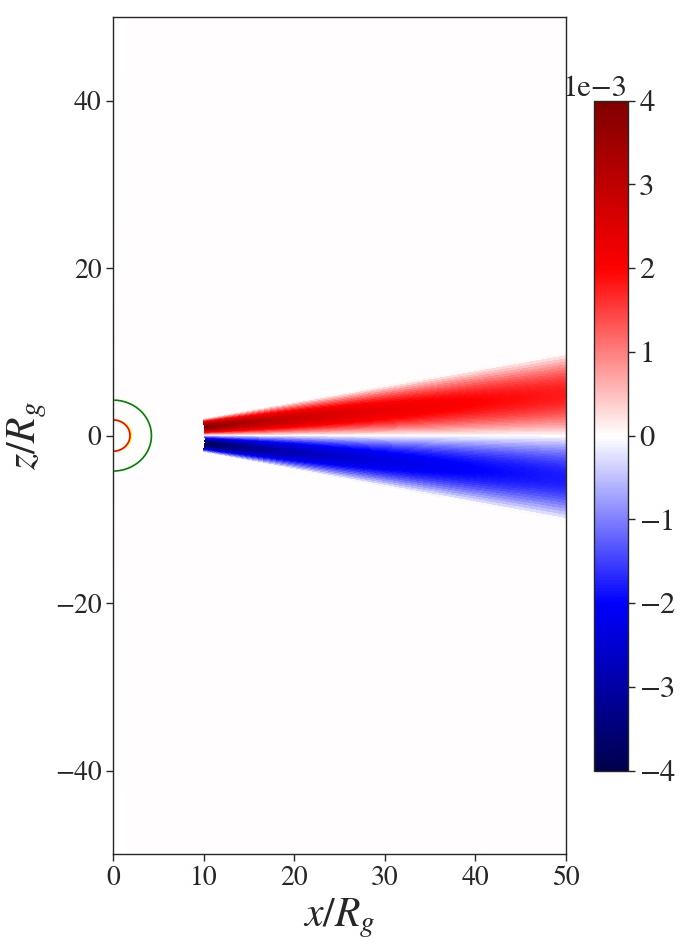}
    \includegraphics[width=0.65\columnwidth]{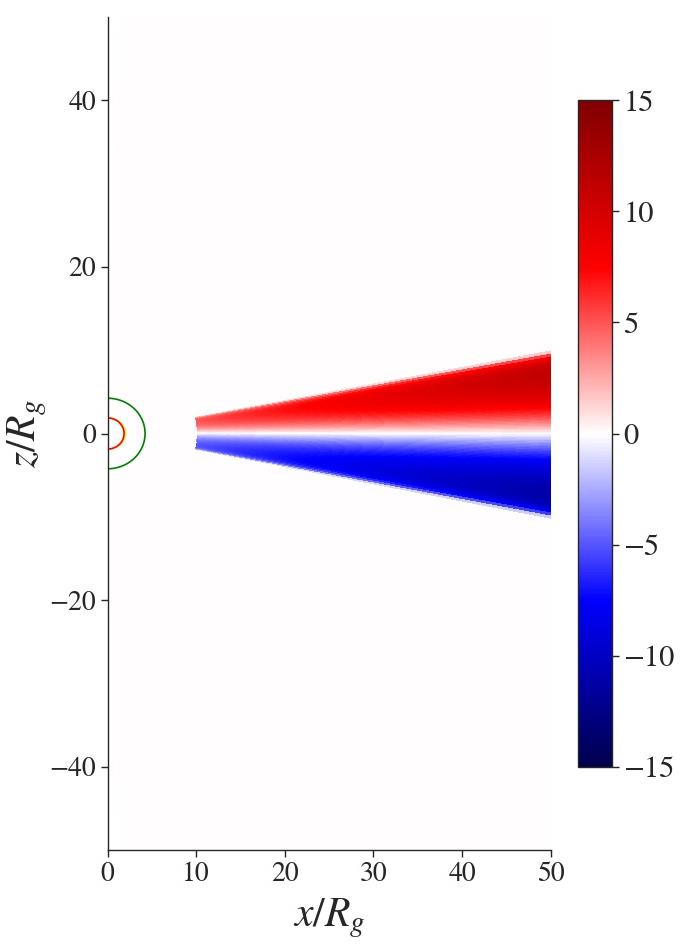}
\caption{Distribution of the magnetic diffusivity $\eta$, the dynamo parameter $\xi$,
and the magnetic Reynolds number ${\cal R}_{\xi}$. Representations are in log scale. }
\label{fig:eta_xi}
\end{figure*}

The simulation is initiated with a thin accretion disk with a density distribution similar to the 
one used by \citet{Vourellis2019}
\begin{equation}
    \rho(r,\theta) = \left[ \frac{\Gamma -1}{\Gamma} \frac{r_{in}}{r} \frac{1}{\epsilon^2} \left( \sin \theta + \epsilon^2 \frac{\Gamma}{\Gamma -1} \right) \right]^{1 / (\Gamma -1)},
    \label{eq:Somayeh}
\end{equation}
which is connected with the gas pressure by an ideal equation of state $P = K\,\rho^{\gamma}$.
The aspect ratio of the disk is $\varepsilon \sim 0.1$.
The disk is surrounded by a corona with an initial density and pressure given by
\begin{equation}
\rho_{\text{cor}} \propto r^{1 / (1-\Gamma)}, \quad\quad p_{\text{cor}} = K_{\text{cor}} \rho_{\text{cor}}^{\Gamma}.
    \label{eq:corona}
\end{equation}
We choose $K=0.001$ for the disk and for the corona $K_{\text{cor}}=1$ and force an initial pressure equilibrium 
between he disk and the corona implying a density jump between corona and disk surface.
When the simulations starts, the initial corona starts collapsing, and is then replaced by the floor values 
for density and pressure, described by a broken power \citep{Vourellis2019}.

The disk is given an initial rotation profile following \citet{PW} as
\begin{equation}
    \Tilde{u}^{\phi} = r^{-3/2} \left(\frac{r}{r-R_{\text{PW}}} \right),
    \label{eq:PW}
\end{equation}
where $R_{\text{PW}}$ is a constant, here set equal to the gravitational radius $\rg$.

The initial seed magnetic field is purely radial and is prescribed only inside initial disk.
Having a radial field has the advantage of a vanishing shear between the rotating disk and disk corona.
We have also run simulations starting with a purely toroidal initial field, however, resulting
in a similar long-term magnetic field evolution.
The initial magnetic field prescription involves purely radial magnetic field lines that converge in the black hole horizon, implemented numerically via the magnetic vector potential, 
\begin{equation}
    A_{\phi}(r,\theta) \propto \exp{ \left[-\frac{1}{2} \left( \frac{1}{0.05 \tan\theta} \right)^2 \right]}.
    \label{eq:VecPot}
\end{equation}
The strength of the seed field is defined by the choice of plasma-$\beta = 10^6$.

In the simulations we apply a profile of the magnetic diffusivity that is somewhat modified compared to
\citet{Vourellis2019}, but similar to the dynamo simulations by \citet{Stepanovs2}.
This profile has a constant plateau across equatorial plane, but then quickly drops in polar direction 
for $\theta < 85^{\circ}$ and $\theta > 95^{\circ}$,
\begin{equation}
    \eta(r,\theta) = \eta_0 \exp \left(-100 \frac{\left(\pi/2 - \theta\right)^4}{\arctan(\chi_{\eta} \cdot \varepsilon)} \right),
    \label{eq:diffusivity}
\end{equation}
where $\eta_0$ is the (initial) maximum value and $\chi_{\eta} = 3$ characterizes the scale height of the 
diffusivity distribution.
The profile of the mean-field dynamo follows \citet{Stepanovs2},
\begin{equation}
    \xi(r,\theta) = \xi_0 \frac{1}{\sqrt{r/r_{\text in}}} \sin\left( \frac{\pi}{\chi_{\xi} \cdot \varepsilon \cdot \tan{\theta}}\right),
\end{equation}
where $\chi_{\xi} = 1$ is a characteristic scale height for the dynamo.
The latter is an important constraint, as by choosing $\chi_{\xi} < \chi_{\eta}$ we ensure that the dynamo 
action is always enclosed in the resistive area.
Otherwise we may consider arbitrarily large dynamo numbers (see below).
Furthermore, the purely resistive outer layers of the disk will allow for easier launching 
and mass loading of disk winds \citep{Vourellis2019}.
In Figure~\ref{fig:eta_xi} we show the distribution of the magnetic diffusivity $\eta$, 
the dynamo parameter $\xi$ and Reynold's number ${\cal R}_{\xi}$.

%=========================================================================================================
\section{The accretion disk dynamo}
\label{sec:tddynamosims}
Our paper focuses on the generation of a strong poloidal flux anchored to a thin accretion disk.
As reference simulation we consider the case of a Schwarzschild black hole.
A very low radial magnetic field with $\beta = 10^6$ is applied as a seed field for the dynamo action,
following the (non-relativistic) approach by \citet{Stepanovs2, FendtGassmann2018}.
Note that our initial disk structure is not in complete equilibrium with the black hole (different from the non-relativistic simulations).
Thus, during the initial evolution the disk will adjust on the relativistic potential,
causing deviations on the seed field from the perfectly radial structure.
These deviations, however, are minimal and significantly smaller that the magnetic field
that is generated by the dynamo later on.
Simulations with a different initial field structure (e.g. a toroidal field) lead to the same
magnetic structure in the saturation phase.

\begin{table}[t]
    \caption{Summary of thin disk simulation runs. 
    Listed is the ID of the run; 
    the Kerr parameter $a$ applied; 
    the maximum magnetic diffusivity $\eta_0$, typically located at the inner disk radius; 
    the maximum (initial) dynamo parameter $\xi_0$; 
    and the quenching parameter $\beta_{\textrm eq}$, corresponding to an equipartition field strength, beyond which dynamo action is strongly quenched.
    }
    
    \centering
    \begin{tabular}{ccccc}
    \hline
    \noalign{\smallskip}
        run   &  $a$  & $\eta_0$ & $\xi_0$ & $\beta_{\textrm eq}$ \\
    \noalign{\smallskip}     
    \hline 
    \noalign{\smallskip}
        sim0    &  0  & 0.001 & 0.004 & 1000 \\
        sim1    &  0  & 0.001 & 0.004 & 100 \\
        sim2    &  0  & 0.001 & 0.004 &  10  \\
        sim3    &  0  & 0.001 & 0.004 &  1   \\
        sim0.1  & 0.9 & 0.001 & 0.004 & 1000 \\
        sim1.1  & 0.9 & 0.001 & 0.004 & 100  \\
        sim4    &  0  & 0.001 &   0   &  -   \\
        sim5    &  0  & 0.001 & 0.004 & no quenching  \\
    \noalign{\smallskip}
    \hline
    \end{tabular}
    \label{tab:simulations}
\end{table}

\begin{figure*}
\centering
    \includegraphics[width=0.67\columnwidth]{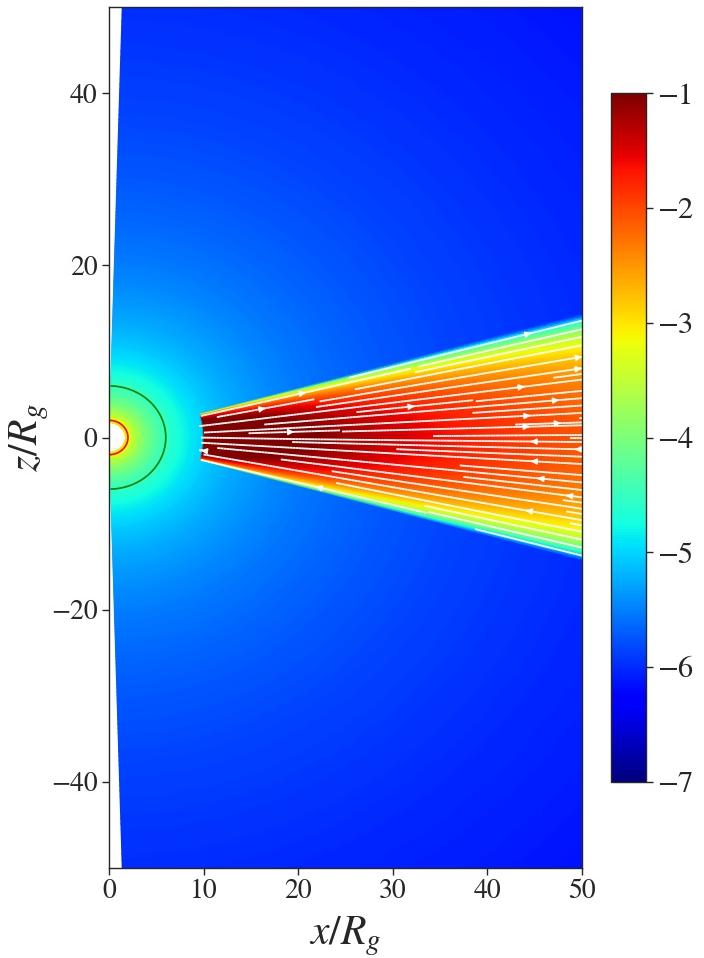}
    \includegraphics[width=0.67\columnwidth]{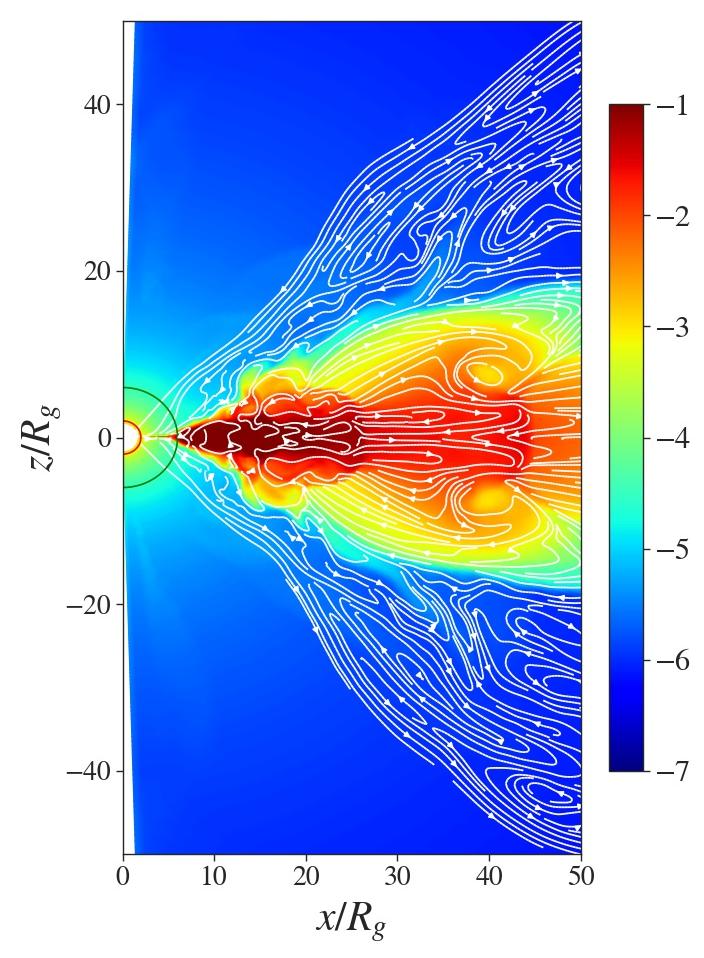}
    \includegraphics[width=0.67\columnwidth]{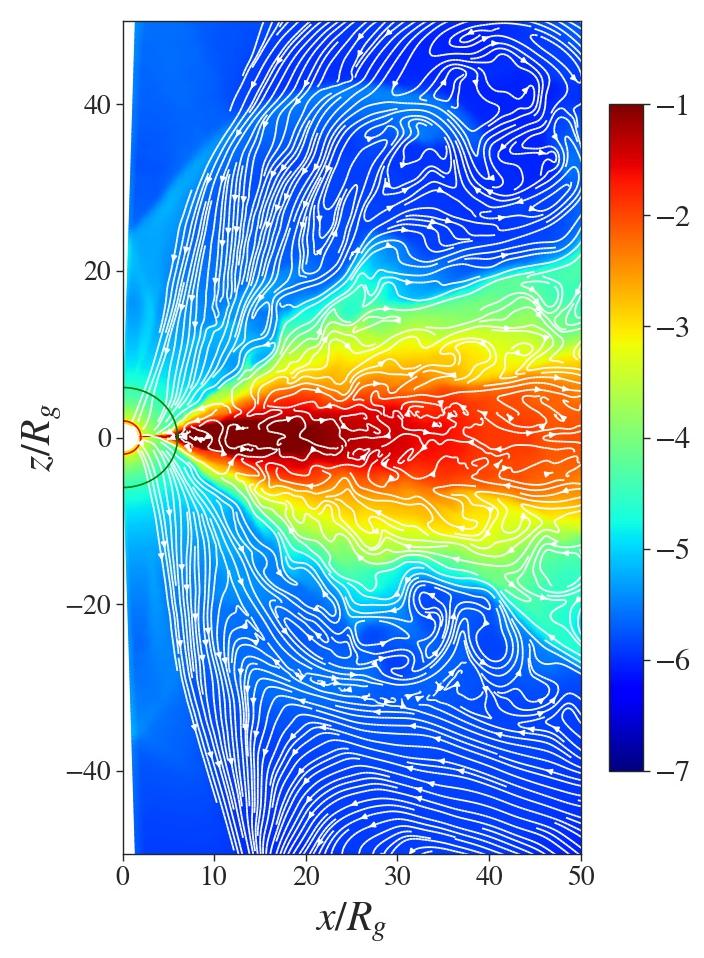}
    \includegraphics[width=0.67\columnwidth]{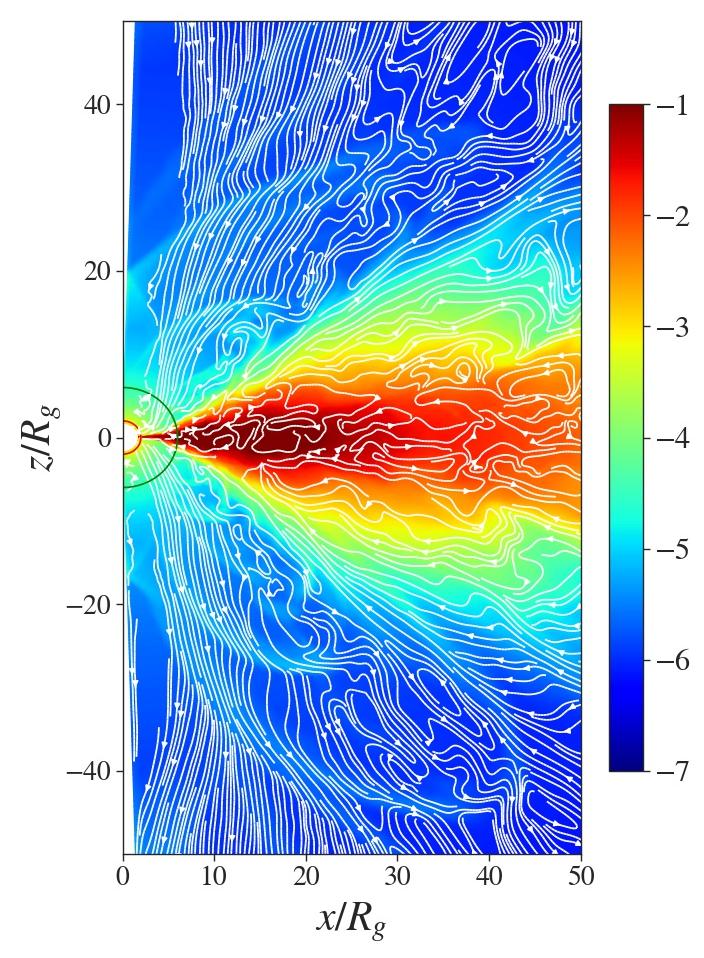}
    \includegraphics[width=0.67\columnwidth]{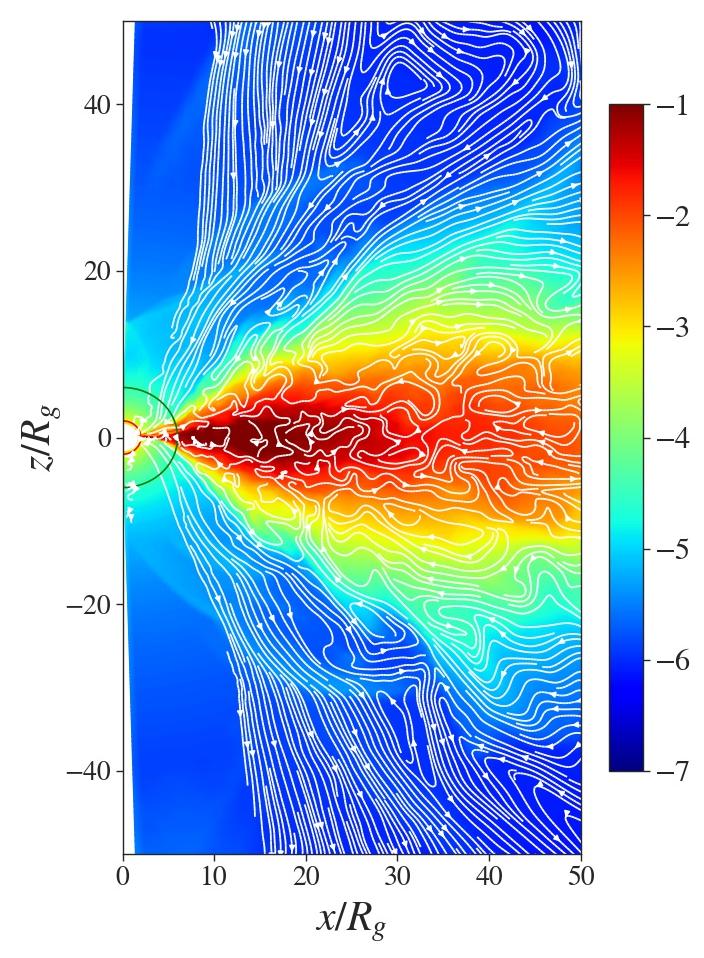}
    \includegraphics[width=0.67\columnwidth]{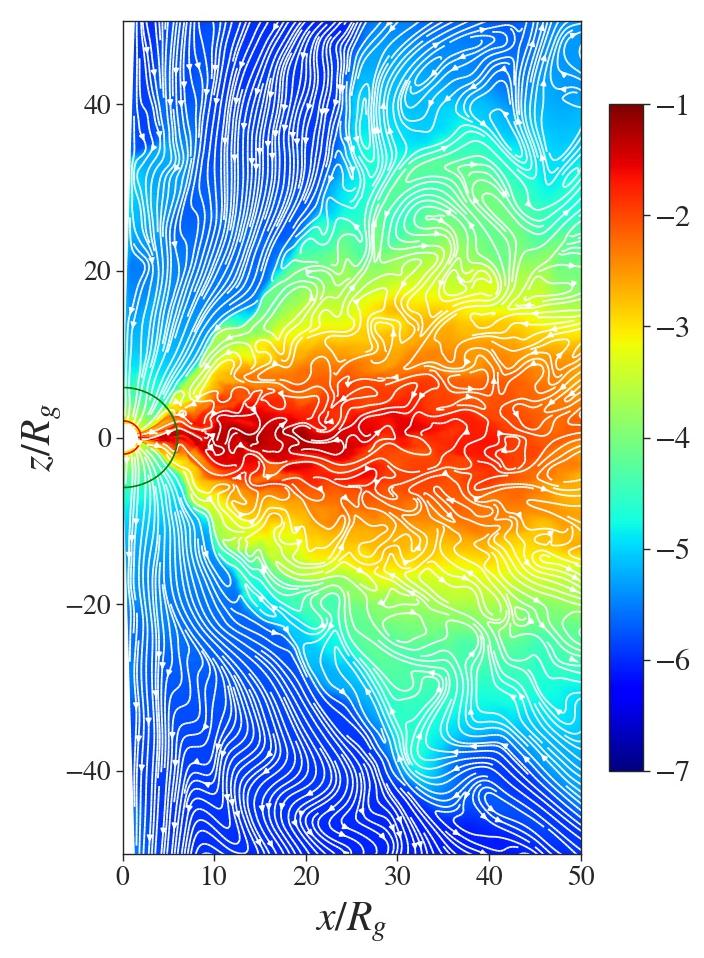}
\caption{Evolution of the disk density and the poloidal magnetic field (white lines) for simulation {\em sim1} at times $t=\left(0, 1000, 3000, 6000, 9000, 12000 \right)$. The red line around the $(0,0)$ point is the event horizon and the green line denotes the radius of the ISCO.} 
\label{fig:evolution1}
\end{figure*}

Figure~\ref{fig:evolution1} shows exemplary evolution of the poloidal magnetic field together 
with the evolution of the density for reference simulation {\em sim1}.
When the simulation starts, the initial radial structure of the magnetic field inside 
the disk is evolved by the dynamo action.
The geometry of this dynamo-generated field does not follow a clear dipolar structure.

In addition to the dynamo action, the hydrodynamic turbulence that develops in the disk 
amplifies the  state of the magnetic field.
We note here that this turbulence is a consequence of the existence of an anti-aligned poloidal
field structure that further develops when the poloidal field is advected towards the horizon.
The field geometry close to the horizon (say for $r<5$) resembles that of the Wald solution 
\citep{1974PhRvD..10.1680W, 2005MNRAS.359..801K}.
With our resistive approach, this leads to strong reconnection events that affect the mass loading
of the outflow as well as the accretion of material, and also the foot point of the magnetic field
lines that guide the outflow (see also \citealt{Vourellis2019}). 

We do not expect the MRI to be detected in our simulations.
In fact our resolution would be almost sufficient to resolve the some of the large wave modes.
So while the MRI might potentially be observed in an ideal MHD study, for resistive MHD the excitation 
of the MRI modes is severely damped.
Here, we refer to \citet{Fleming2000} and \citet{LongarettiLesur2010} who investigated in great detail
the effect of resistivity on the evolution of the MRI.
\citet{QQ1}, investigating the role of the MRI in a general relativistic torus in resistive MHD 
came to the conclusion that the MRI is more and more damped for increasing level of resistivity
(although applying a lower resolution).

As soon as the dynamo-induced magnetic field is strong enough to remove angular momentum
from the disk, accretion sets in.
With accretion going on, the newly generated field is advected towards the black hole.
In the last stage of the simulation the environment close the black 
hole is strongly magnetized by a poloidal field with a dipolar configuration.
However, at larger radii and especially inside the disk, at the time scales investigated, the dynamo has not yet generated a clear structure that provides a large scale magnetic flux.
Note that these simulations are quite CPU-intensive, and the time scales we are reaching here
of about $t=10000$ corresponds to about 150 inner disk rotations. 
In contrast, the non-relativistic disk dynamo simulations 
\citep{Stepanovs2, FendtGassmann2018}
reach up to several 100.000 disk rotations.

\begin{figure}
\centering
    \includegraphics[width=0.98\columnwidth]{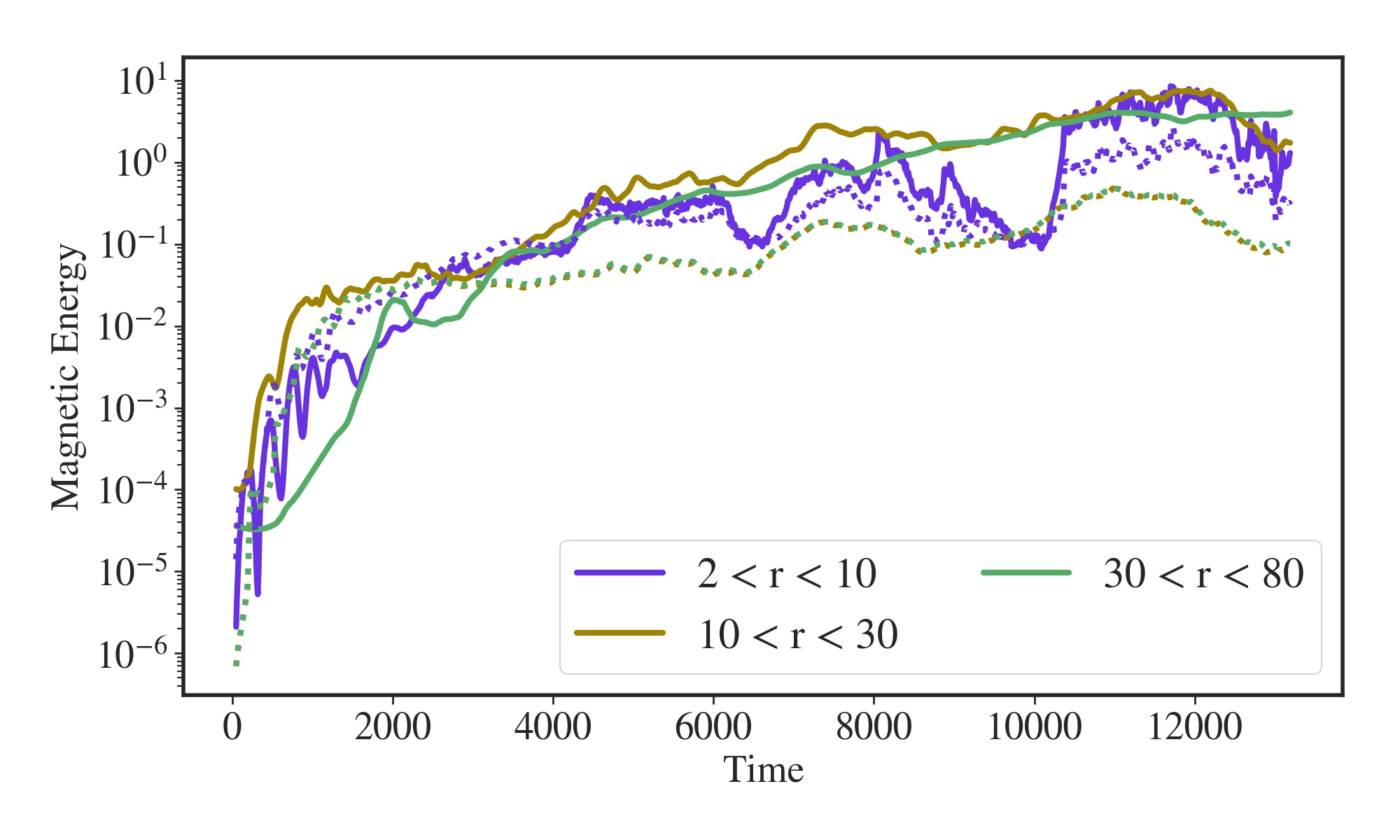}
\caption{Evolution of the poloidal (solid line) and toroidal (dotted line) magnetic energy integrated between 
three different radii inside the accretion disk for simulation {\em sim1}. }
\label{fig:magnener}
\end{figure}

%---------------------------------------------------------------------------------
\subsection{The evolution of the dynamo generated magnetic field structure}
We now discuss the magnetic field structure that is generated by the disk dynamo (see Figure~\ref{fig:evolution1}) and how
its geometry and field strength depends on the parameters that trigger the dynamo process, 
namely the strength of the dynamo $\xi$, the disk diffusivity $\eta$ (in combination the dynamo number), and the 
quenching parameter $\mu_{\rm eq}$.

Figure~\ref{fig:magnener} shows the evolution of the integrated poloidal and toroidal magnetic energy 
in three different regions (distinguished by their inner and outer radius).
We emphasize that for later times the dynamo induced poloidal field remains substantially higher than the toroidal field
component.
This again points towards an efficiently working $\alpha$-dynamo, since only the $\alpha$-process can amplify the poloidal field 
component.

\begin{figure}
\centering
    \includegraphics[width=0.98\columnwidth]{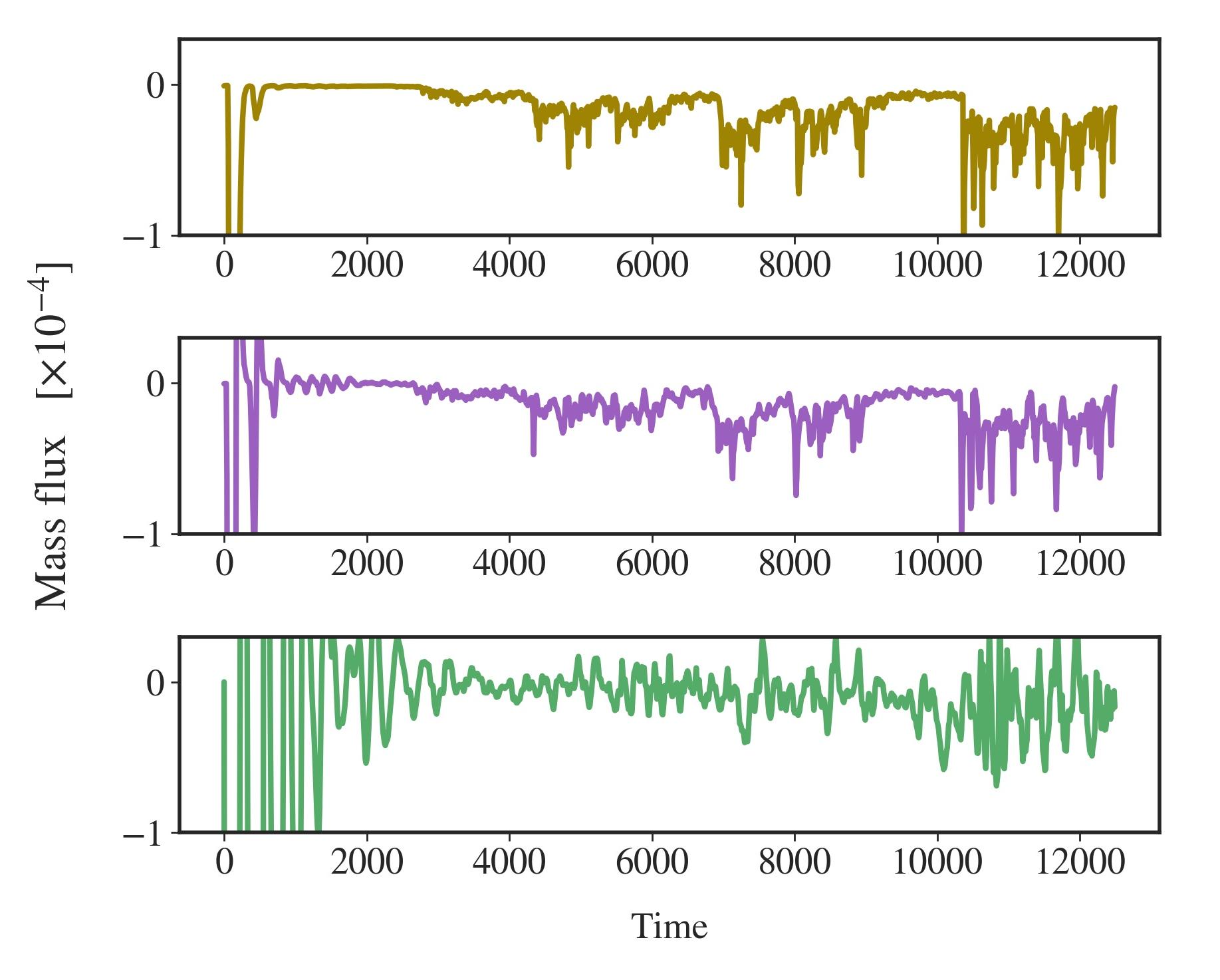}
    \includegraphics[width=0.98\columnwidth]{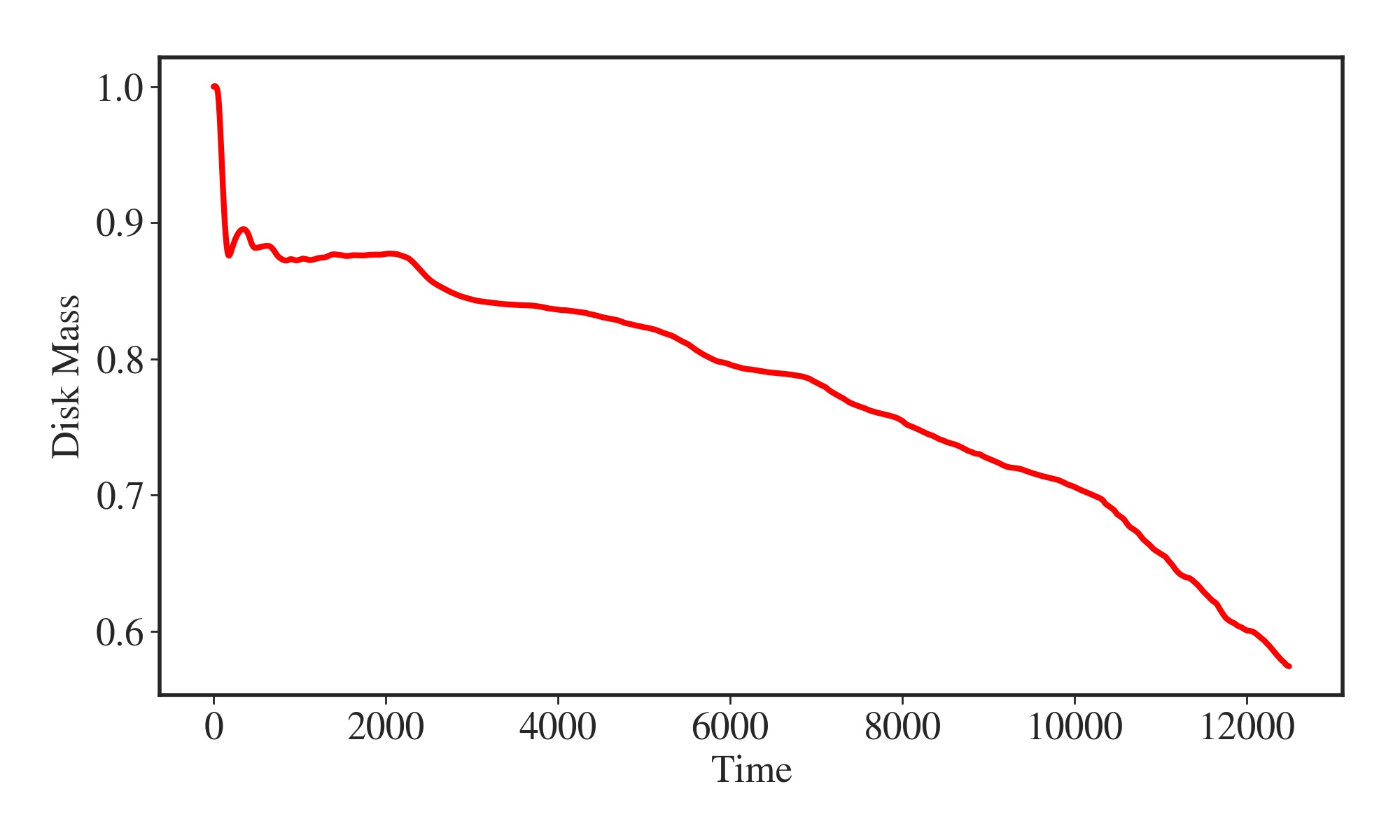}
\caption{Evolution of mass flux rate measured at at $r \sim (3, 6, 11)$ (top) and normalized disk mass (bottom) for polar
angles $75^\circ < \theta < 105^\circ$ for simulation {\em sim1}. 
The negative values denote the accretion rate.
The mass flux is also normalized with the initial disk mass.}
\label{fig:accretionrate}
\end{figure}

For the innermost part of the accretion stream between the black hole horizon and the initial inner disk radius $(2<r<10)$, 
the initial magnetic energy vanishes (since the seed field is confined inside the disk only) and 
then quickly increases due to the dynamo-generated field.
When the disk starts evolving, meaning that accretion of disk material sets in, the magnetic field is getting advected 
towards the black hole.
Due to the unsteady accretion process, also the advected magnetic energy is varying.
Indeed, the initial variations in magnetic energy we find, do correlate with the spikes in the accretion rate 
(see top panel of Figure~\ref{fig:accretionrate}).
After $t \simeq 2000$, the increase of the magnetic energy continues, but with a smaller slope and 
after $t \simeq 6000$ we see indication of a saturation of the dynamo action as the slope of the magnetic energy increase
decreases even further to the point it is almost constant.
Still, we do not reach a constant value, as at $t \simeq 10000$ there is another sudden increase in the magnetic energy
that is coinciding with another strong accretion event.

At larger disk radii the time evolution of the magnetic energy looks somewhat different.
The initial variations do not appear, as the accretion process works smoother in these areas.
However, the overall behaviour of the magnetic energy is similar to the innermost disk.
Again, at the beginning the energy increases much faster for the middle part $(10<r<30)$ but after $t \simeq 3000$ the disk magnetic 
energy is almost the same for both regions and slightly larger than the energy in the innermost part.
After $t \simeq 10000$, with the sudden energy increase from the inner part, the energy in all three radii is approximately the same.
The comparison of the magnetic energy content in these disk areas of course depends on the choice of the integration areas.
What we learn from this comparison is that the disk dynamo works similarly efficient over almost the whole disk.
Further, the slope of the magnetic energy growth is comparable between these areas, indicating a well posed setup for the disk dynamo,
in particular, a direct coupling between the processes of accretion, ejection and dynamo action.

\begin{figure}
\centering
        \includegraphics[width=0.98\columnwidth]{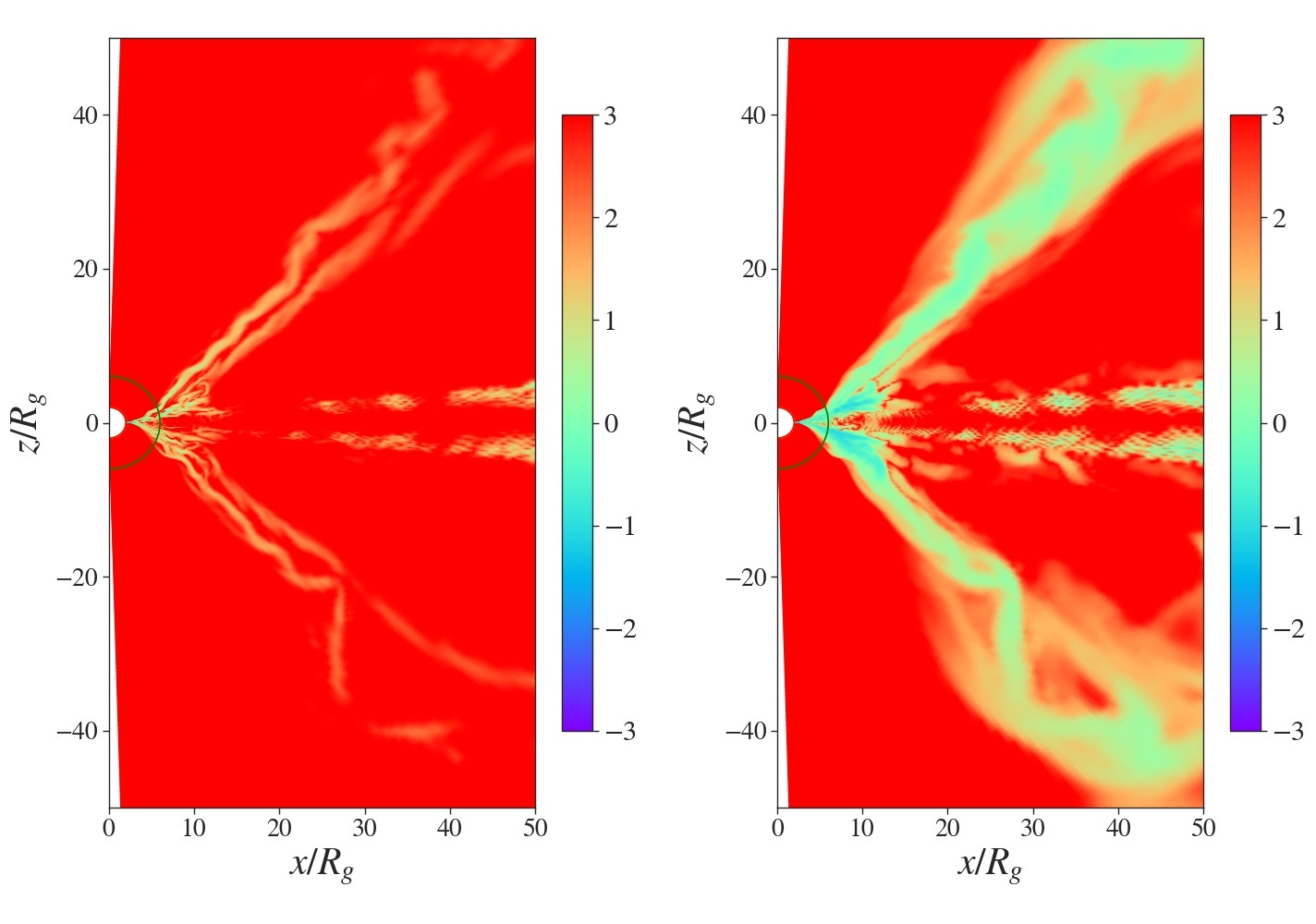}
        \includegraphics[width=0.98\columnwidth]{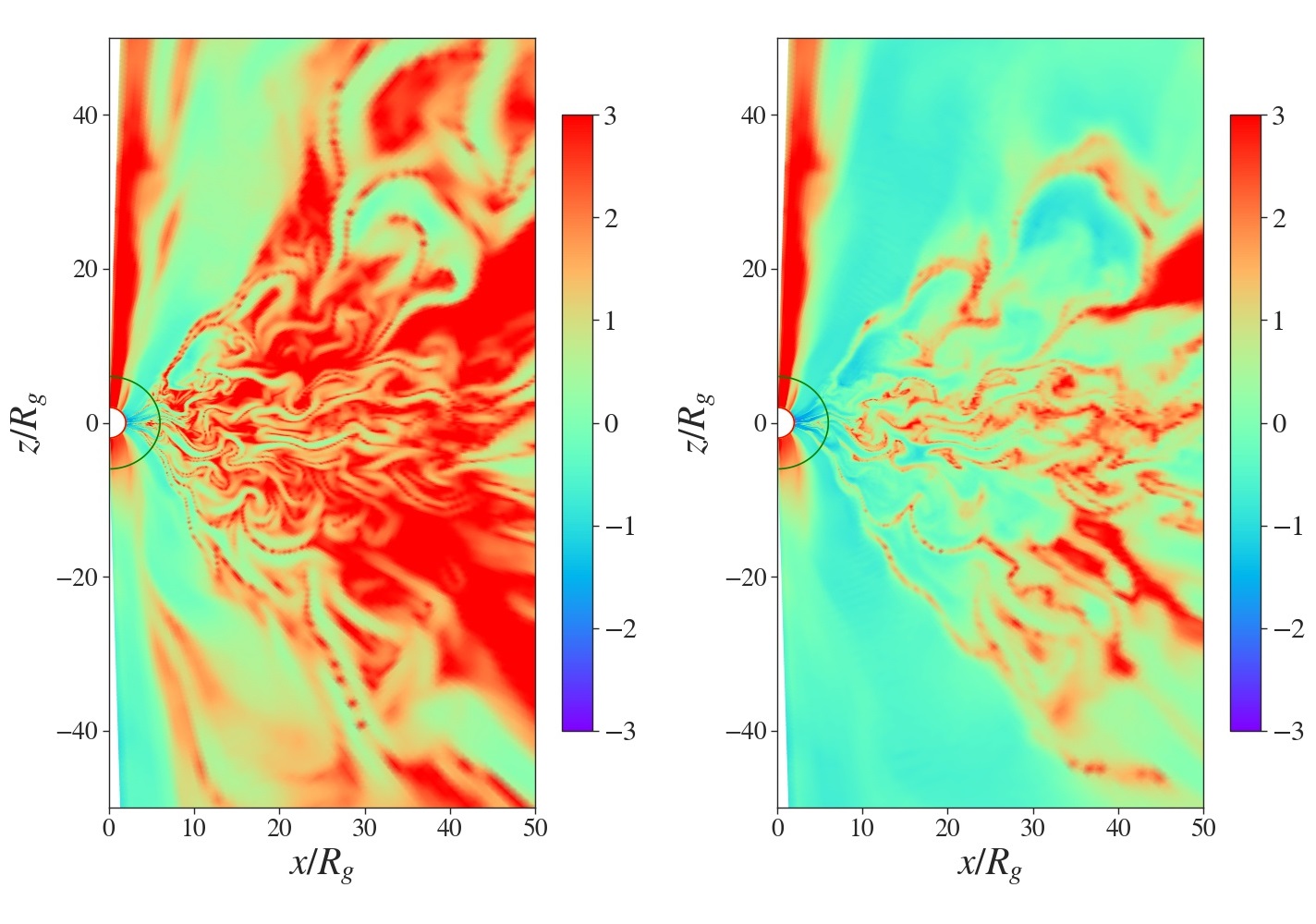}
\caption{Plasma-$\beta$ at simulation times $t=4370$ (top) and $t=12000$ (bottom),
considering the poloidal (left) and the toroidal (right) magnetic field component,
respectively, for simulation {\em sim1}.}
\label{fig:magnbeta}
\end{figure}

Figure~\ref{fig:magnbeta} shows distribution of the plasma-$\beta$ for the poloidal and the toroidal magnetic field components at
two different time steps for simulation {\em sim1}.
Obviously, due to the dynamo activity, the plasma-$\beta$ decreases with time - first in inner disk area and later also in the
main disk body.
Also the disk wind that is launched at later time carries a low plasma beta.
Overall, the toroidal magnetic field component is dominating. 
This has been observed also in \citet{Vourellis2019} for a magnetic field that is initially prescribed 
and is consistent with the literature of GRMHD disk simulations.
For the time scales considered here, this tells us that the differential rotation ($\Omega$ effect) still dominates the 
turbulent dynamo ($\alpha$ effect).
At $r\sim 15$ along the equatorial plane the dynamo number characterizing differential rotation is
${\cal R}_{\Omega} \approx 59$,
while the maximum $\xi$ dynamo number ${\cal R}_{\xi}$ does not exceed 13 (see Eq. \eqref{eq:dynamonumber}).

However, we see that the plasma-$\beta$ measured in the inner part of the disk and in the accretion funnel do slightly increase during the 
later times of the simulation.
This is a direct effect of the quenching mechanism we implemented.
In addition, the magnetic field becomes quite entangled in the later stages of the simulation (see last panels of
Figure~\ref{fig:evolution1}).
As a consequence, there exist further effects that can lead in the dissipation of the magnetic field, in particular magnetic reconnection.
What can be clearly seen from the figure is that exactly those areas of low plasma-$\beta$ are also the areas from which strong outflows
develop from the inner disk (see also Section~\ref{sec:diskwind} below).

\begin{figure}
\centering
    \includegraphics[width=0.98\columnwidth]{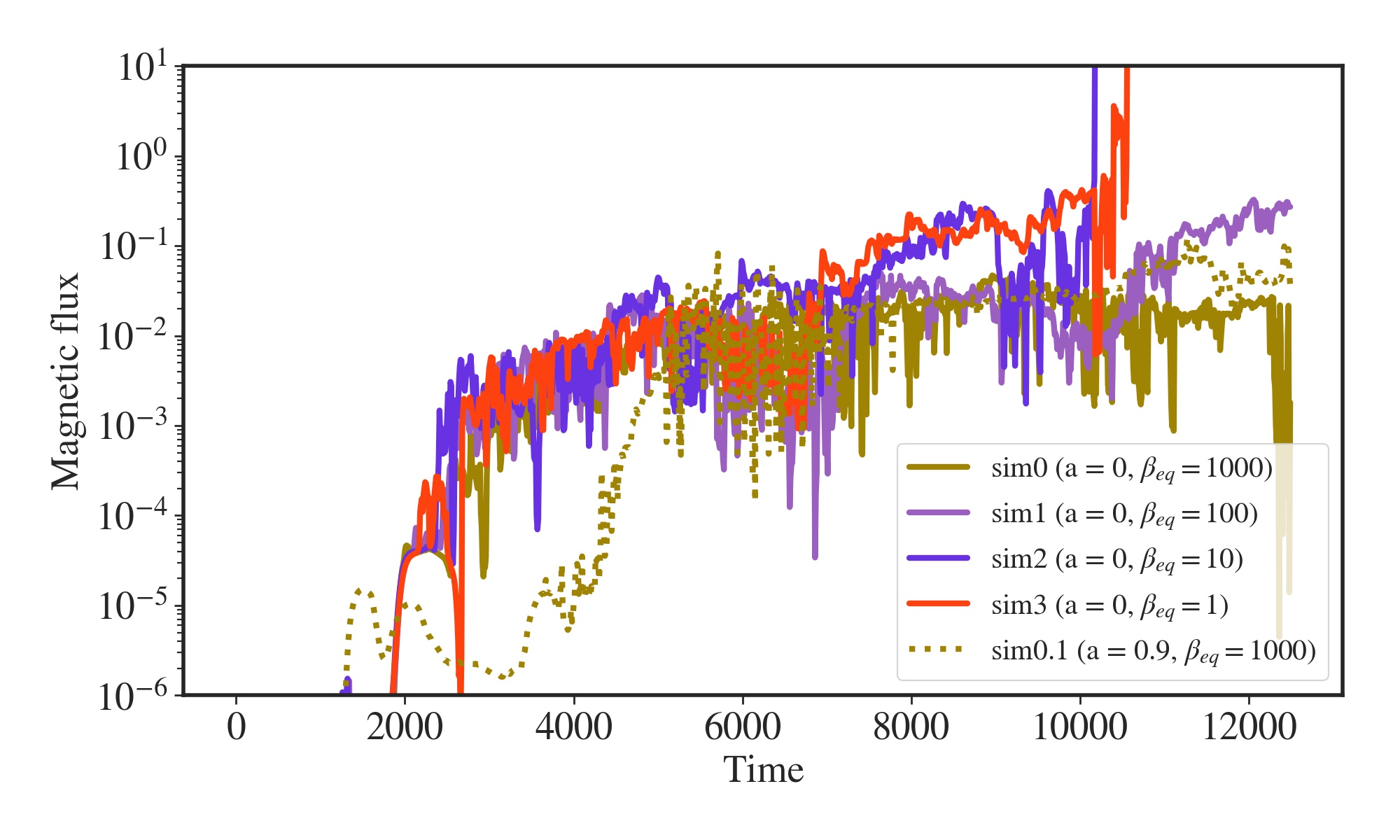}
    \includegraphics[width=0.98\columnwidth]{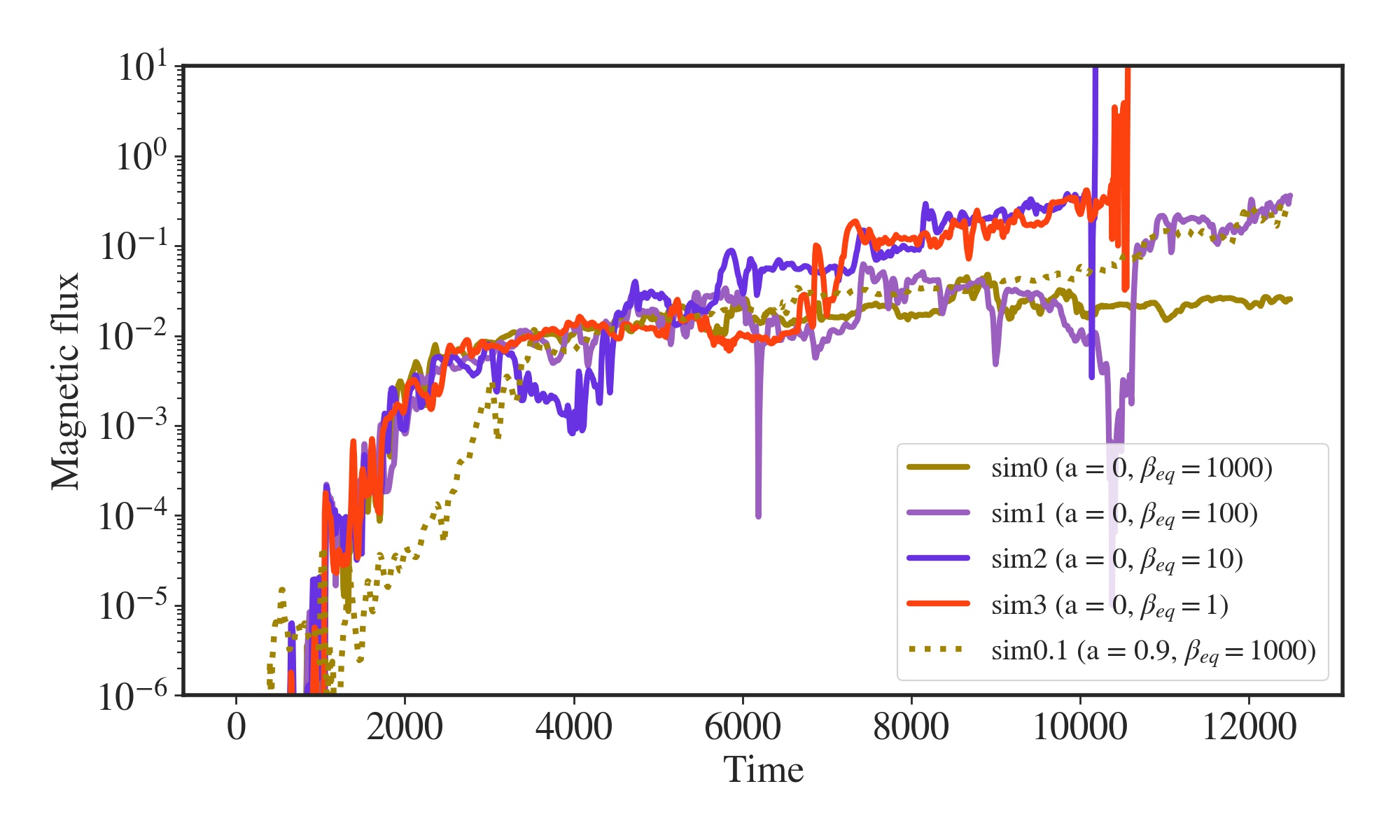}
\caption{Time evolution of the absolute integrated magnetic fluxes. 
Shown is the magnetic flux from the disk surface (top), thus integrated radially along the disk surface,
and the flux through a circular cross section over the polar angle $0^{\circ}<\theta<70^{\circ}$ 
at radius $r \sim 11$ and its corresponding areas in the lower hemisphere (bottom).
Simulations with different levels of dynamo quenching and Kerr parameter are considered.
}
\label{fig:magneticflux_absB}
\end{figure}

An essential condition for jet launching is a sufficient {\em magnetic flux} of the jet source, synonymous for a large-scale 
open magnetic field structure.
As already briefly mentioned, although we may measure a substantial (poloidal) magnetic energy $\propto B_{\rm p}^2$ 
in the disk, the poloidal magnetic flux $\int \vec{B}_{\rm p} d\vec{A}$ may vanish on average, if the field is strongly tangled.

It is therefore interesting to see how the magnetic flux evolves that is generated by our disk dynamo.
This is shown in Figure~\ref{fig:magneticflux_absB} for simulations with different Kerr parameters and quenching thresholds.
Here we choose two different representations of the absolute integrated magnetic flux.
One option is to integrate along a circle of constant radius (lower panel) and 
the other is to integrate along the disk surface (upper panel),
using the $B_r$ and  $B_{\theta}$ components of the magnetic field, respectively.
Note that radius of the circle of $r=11$ is chosen such that it measures the magnetic flux close to
the black hole which potentially may launch a Blandford-Znajek jet.
The angle for the integration is chosen so that we avoid to account for flux from inside the disk where 
the field is more entangled.

The increase in magnetic flux is similar to the time evolution of the magnetic energy 
(see Figure~\ref{fig:magnener}).
Differences are due to the fact that when integrating the cumulative magnetic flux $|\int \vec{B}_{\rm p} d\vec{A}|$,
we effectively average over fluxes in opposite directions, as discussed above.

This clearly demonstrates that our disk dynamo also generates a substantial magnetic flux when generating magnetic energy
(see Figure~\ref{fig:evolution1} showing indeed large-scale open poloidal field lines).
Interestingly, we find that all dynamo simulations evolve with an initially very similar behaviour.
At the later evolutionary stages they, however, diverge, due to the different quenching threshold they obey.
Simulation {\em sim0} keeps an almost constant magnetic flux during its final part, 
while for simulations with a higher quenching threshold the flux increases with a steeper slope. 
Simulations {\em sim2} and {\em sim3} end earlier due to the higher magnetization levels that is allowed by the quenching.
Simulation {\em sim0.1} that considers a highly spinning black hole, also provides an ever increasing magnetic flux 
over the simulation time, however, the dynamo effect is somewhat delayed.

For comparison, we have also performed test simulations {\em sim4} without any dynamo action 
and {\em sim5} without quenching.
As expected, in simulation {\em sim4} the initial radial field structure is conserved while
slightly expanding in the surrounding disk corona.
Minor weak outflows can be detected from the disk surface that can be attributed mostly to 
the local force balance between the disk vertical pressure gradient (magnetic pressure is negligible here) and the 
vertical gravitational force (induced by the black holes metric).
In simulation {\em sim5} the magnetic field develops as in simulation {\em sim3}, however, 
the lack of a quenching mechanism generates a magnetic field that becomes extremely strong and 
that leads the code to crash rather early.
A similar behavior was also observed by \citet{Tomei2020MNRAS.491.2346T}.
Both simulations, {\em sim4} and {\em sim5} may serve as reference for further test simulations of our
implementation of the dynamo and the dynamo quenching.

%-----------------------------------------------------------------
\subsection{Disk evolution - magnetic field and mass fluxes}
We now investigate the 
evolution of the accretion disk, in particular the mass accretion rates, in more detail.
Figure~\ref{fig:accretionrate} (top) show the normalized mass flux through surfaces of constant radius $r=(3,6,11)$.
We have selected a slice between $75^{\circ} < \theta < 105^{\circ}$ in order to account for 
the initial disk structure and the evolution of the disk.

Due to the weak seed field, the initial evolution is basically hydrodynamic.
While the initial radial structure of the magnetic field is ideal for angular momentum removal from the disk material,
its strength is too low in order to have a big impact.

We find that right in the beginning of the simulation, the accretion rate shows large variations.
This is due to the lack of a perfect vertical hydrodynamic equilibrium in the disk.
This leads to distortions in the disk structure, and to episodic accretion events of material that has moved inside the 
ISCO (note the initial position of the inner disk radius is at $r=10$, well outside the ISCO).
The inner part of the disk is quickly advected and when it crosses the marginally stable orbit it disconnects from the 
disk and falls in the black hole leaving a \enquote{gap} behind it which quickly fills with material from the remaining 
inner part of the disk.
The initial radial shape of the initial magnetic field supports the advection of disk material towards the black hole,
however, as mentioned above, this field is not dynamically important.
On the other hand, the motion parallel to the poloidal field also implies that no magnetic flux is advected initially. 

The episodic accretion is repeated several times depending on the disk radius (see the accretion spikes in 
Figure~\ref{fig:accretionrate}, top).
Further out along the disk, the accretion peaks correspond to disruptions that appear at the disk surface 
displacing disk material (see Figure~\ref{fig:evolution1}, top middle panel).
This is a common consequence of the dynamics of the this disk in the relativistic environment \citep{Vourellis2019}.

Around time $t \simeq 2000$ the disk structure settles just outside the marginally stable orbit resulting in a more 
continuous accretion rate.
At this time, the initial seed magnetic field starts being modified by the dynamo leading to initial weak outflows launched 
from the accretion funnel between the marginally stable orbit and the black hole.
The outflows drag the magnetic flux with them, developing a vertical magnetic field component which then contributes to 
establish a smoother accretion rate.
We will later discuss the onset and evolution of these outflows in more detail (see Sect.~\ref{sec:diskwind}).

We find the strongest accretion rates after $t \sim 10000$.
We understand this is essentially caused by the interplay between the dynamo activity and dynamo quenching 
(see Section~\ref{sec:quenching2}).
The weak outflows mentioned above remain present also during the second evolutionary phase,
however, showing a time variation in velocity and mass flux.
Within the accretion disk the magnetic field is substantially entangled, forming loops that reconnect with each other.
The disk starts expanding in the vertical direction, but not yet developing any strong disk wind.

After $t \sim 10000$ we notice an increase in the accretion rate that coincides with an increase in the magnetic field 
strength close to the black hole. 
This is the point in time when the dynamo quenching is not sufficient to stop the magnetic field from increasing further
mainly due to the radial limits within which it is defined.
Note that the dynamo action and its quenching is defined only for $r>10$, which is located inside the accretion disk.
When the magnetic field is advected, the magnetic flux in the black hole magnetosphere increases along with the magnetization, 
but this area is neither dynamo-active nor affected by quenching.
The strong field of the black hole magnetosphere results in divergences in the integration of the equations which 
eventually lead to the termination of the simulation.

When the disk field becomes advected, these field lines also populate the area along the rotational axis resulting in a strong axial
magnetic field that could potentially lead into a Blandford-Znajek-driven outflow 
(in the case of a rotating black hole, see Sect.~\ref{sec:BZjet}).

\begin{figure}
\centering
    \includegraphics[width=0.98\columnwidth]{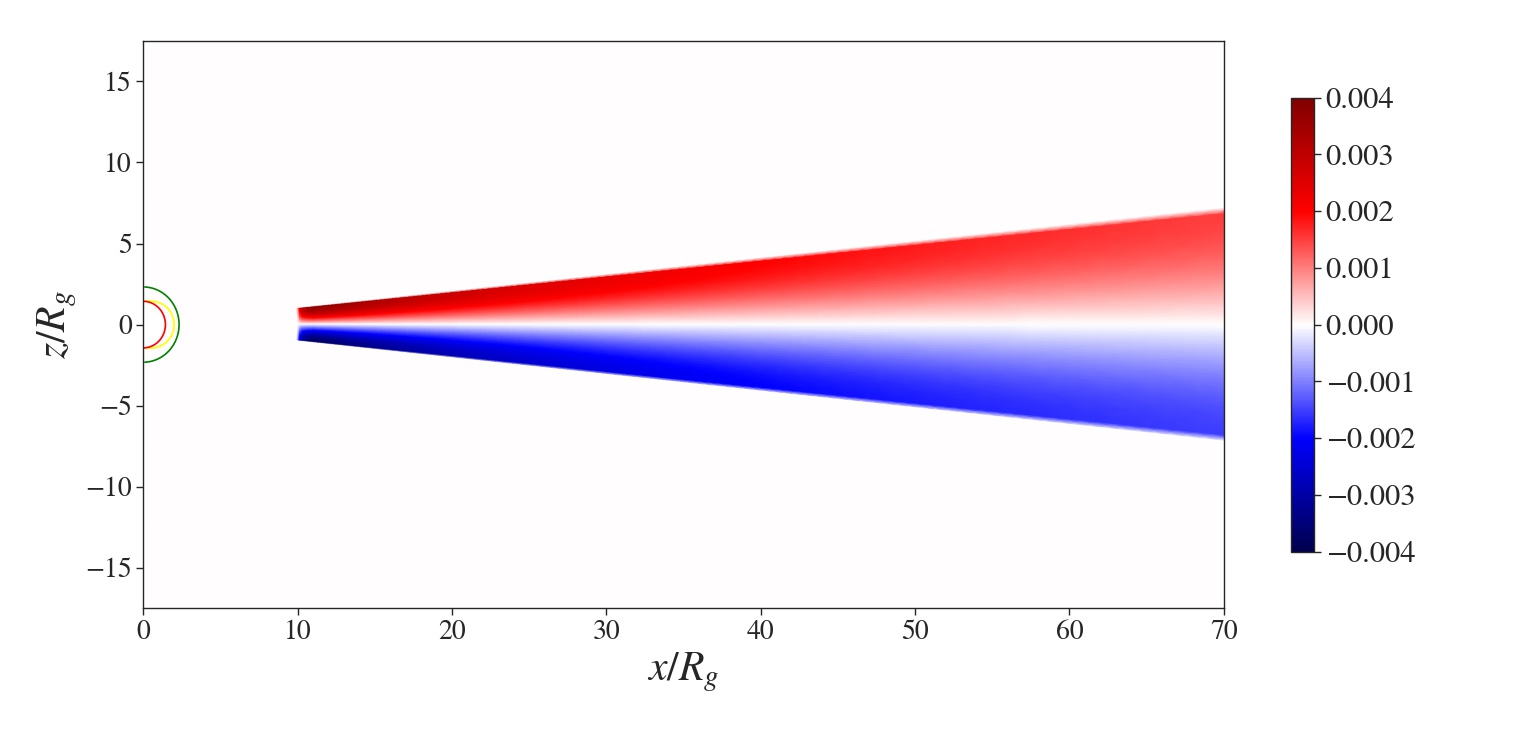}
    \includegraphics[width=0.98\columnwidth]{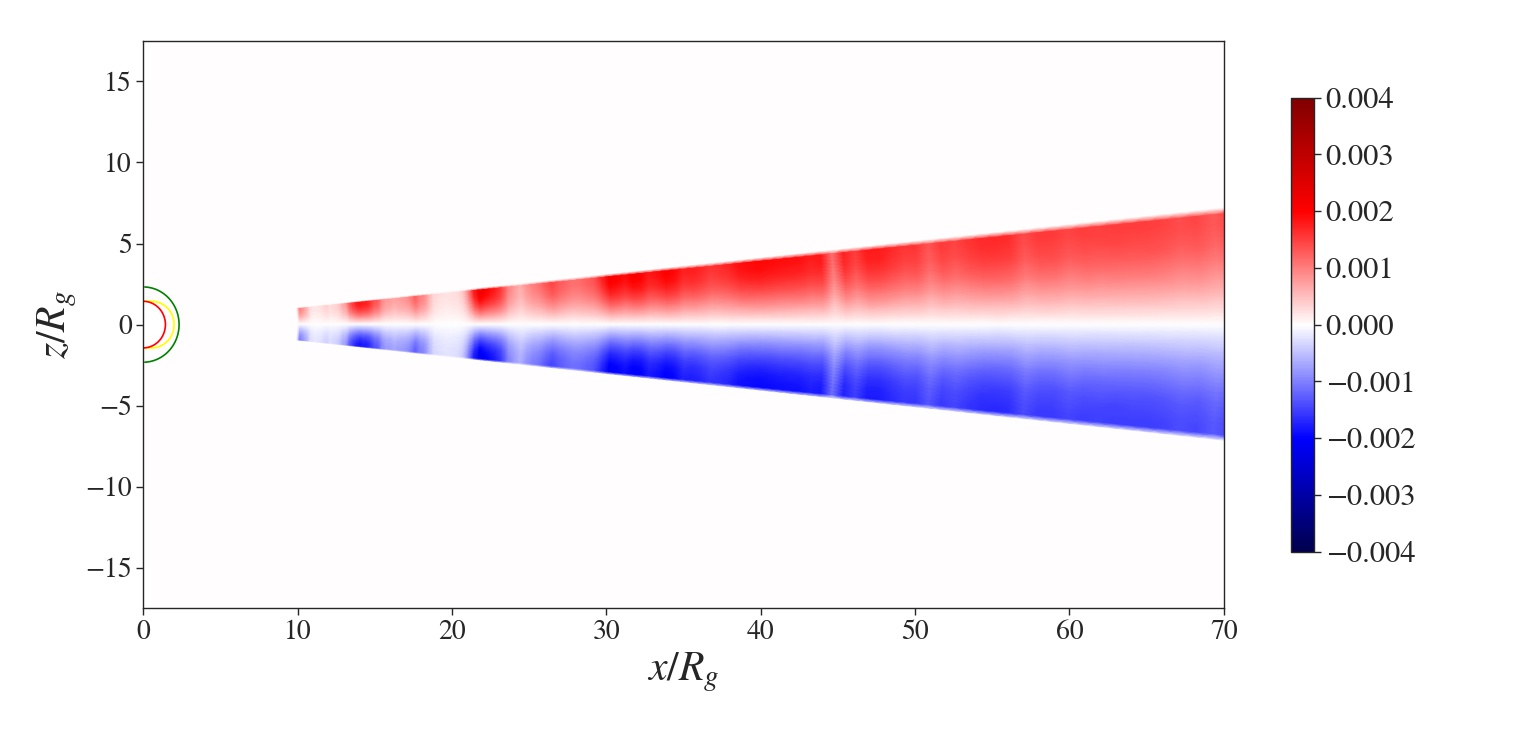}
\caption{The distribution of the dynamo parameter $\xi$ and its variation due to quenching for simulation 
{\em sim0.1} at times $t=1000, 12000$.}
\label{fig:quenching}
\end{figure}

%----------------------------------------------------------------------------------------------------------
\subsection{Dynamo quenching}
\label{sec:quenching2}
For completeness we briefly show the evolution of the dynamo quenching.
After an initial evolutionary phase, the magnetic field continues to increase in both the poloidal and the toroidal component.
However, the toroidal component is undergoing an extra amplification due to the differential disk rotation 
(the $\Omega$-effect).
This later evolution is not as rapid as the initial boost due to the dynamo quenching that starts mitigating the dynamo effect.
The dynamo quenching follows Equation~\eqref{eq:ch3:quenching} which for simulation {\em sim1} has 
$\mu_{\textrm eq} = 1/\beta_{\textrm eq} = 0.01 $.
This implies that as the plasma-$\beta$ inside the disk decreases from the initial value of $10^6$ to an actual value of 100, 
the quenching becomes stronger.

Even for a plasma-$\beta = 1000$, in our parameter setup the dynamo action is already quenched by $\sim 11\%$ and by the time when
plasma-$\beta = 100$ the dynamo parameter will be half of its initial value.
Note that the dynamo quenching works locally, defined by the local vertically averaged disk magnetization.
Depending on this, the actual strength of the dynamo tensor is quenched.

This is shown in Figure~\ref{fig:quenching} where we show the distribution of the dynamo-$\xi$ for two exemplary time steps for simulation
run {\em sim0.1} with a rapidly rotation black hole.
While at $t=1000$ the dynamo still works at its initial strength, at $t=12,000$ quenching is clearly seen at certain disk areas.
The quenching is locally different, implying that the dynamo works with different strength at different positions along the disk and,
while it smoothens the exponential amplification of the field, at the same time it restricts a stronger accretion rate.

%---------------------------------------------------------------------------------------
\subsection{Generating a disk wind}
\label{sec:diskwind}
It is well known that a substantial disk magnetic field is essential for driving a strong disk wind
\citep{BP1982, CK2002, Ferreira1997}.
Also the field inclination plays a role \citep{BP1982}.
In contrast to simulations in the literature considering jet launching from magnetized disk that either 
apply a pre-defined disk field
\citep{Zanni2007, Somayeh2012, Stepanovs3} or self-generate the disk field from a non-relativistic 
dynamo \citep{Stepanovs2, FendtGassmann2018},
in our GRMHD dynamo simulations the resultingdisk  magnetic field does not have the smooth
large-scale structure that may support a Blandford-Payne disk wind due to the symmetry
of the initial field.

\begin{figure}
\centering
    \includegraphics[width=0.48\columnwidth]{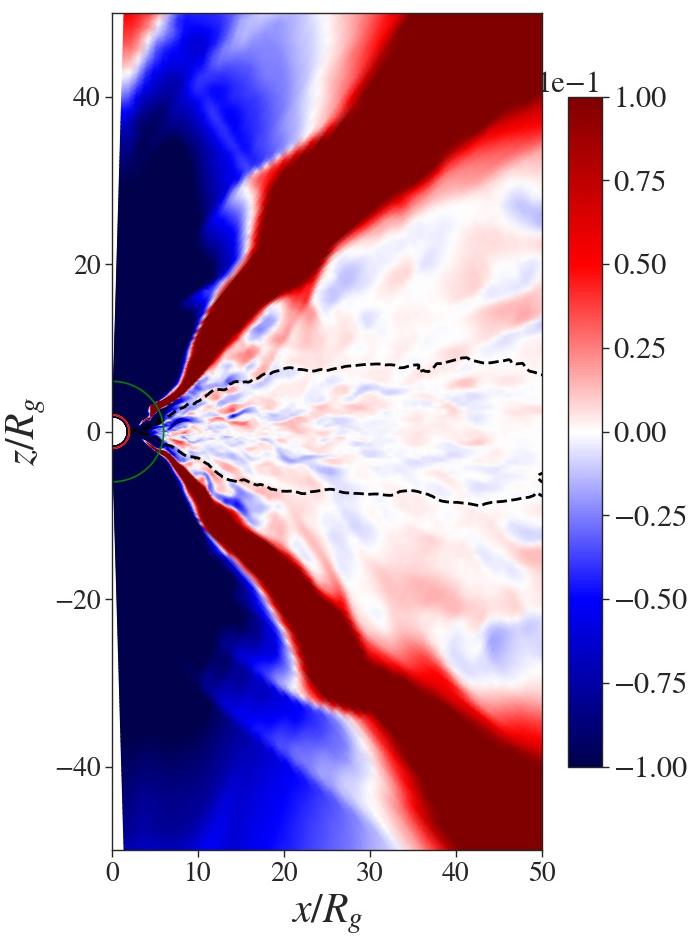}
    \includegraphics[width=0.48\columnwidth]{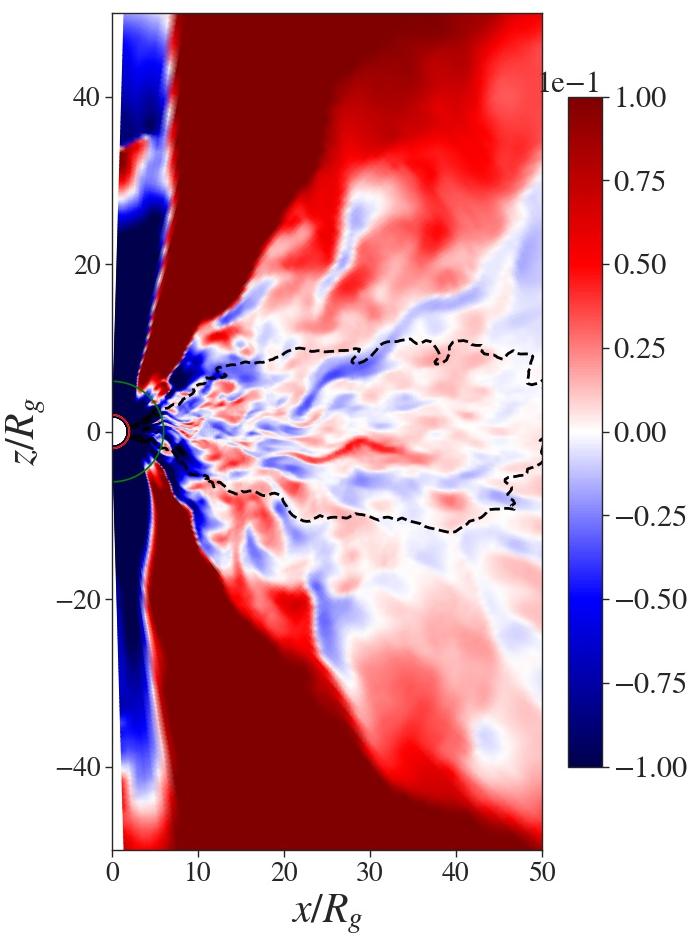}
\caption{Radial velocity of the disk wind in the reference simulation {\em sim1} at times $t = 8000 \text{ and } 12000$.
The black dashed line follows the density contour $\rho \approx 0.003$ which roughly indicates the size of disk and
the transition to the coronal region.}
\label{fig:velocity1}
\end{figure}

However, we can still detect the launching of strong outflows.
From our reference simulation {\em sim1} (see Figure~\ref{fig:evolution1}) we can see the evolution of small scale outflows 
from the disk surface which overall merge into a broad disk wind.
The initially weak winds are launched by a combination of the increased magnetic field pressure gradient
(mainly, but not only, from the toroidal component)
and the re-arrangement of the accretion disk structure in the gravitational field of the black hole 
(advection of magnetic flux).

In Figure~\ref{fig:velocity1} we show the distribution of the radial velocity for the reference simulation at 
times $t = 8000 \text{ and } 12000$.
We also display the density contour $\rho \approx 0.003$ that could be understood as a proxy for the disk 
surface\footnote{As the disk evolves constantly, this choice is somewhat arbitrary. Still the location of the
disk surface is essential to study of the disk-outflow interrelation, in particular the mass loading of the wind.}. 
A better choice would be transition from inflow to outflow, however, for most parts of the disk corona close to the 
disk, there is large-sized area with a constantly positive outwards velocity.
Instead we find patches of both out-flowing and infalling material with velocities somewhat below $(<0.02c)$.
This is the clear signature of a turbulent outflow structure.
We note that here we are still dealing with the initial stages of outflow from the inner parts of 
the disk main body.
Note that the magnetic field and the disk accretion is still evolving, as time $t=10000$ corresponds to only 
50 inner disk orbits, much less than ones achieved in non-relativistic simulations.

An exception is a high-speed outflow that is rooted in the accretion funnel between the black hole and the ISCO.
Here the gas into which the magnetic field is frozen-in orbits with high speed, resulting in radial outflow 
velocities up to $0.2c$.

In the last evolutionary stages the picture changes, however.
The disk wind is now dominated by out-flowing patches moving with $\simeq 0.05c$.
The infalling patches have almost disappeared and we see a substantial disk wind.

In the polar area above and below the black hole we see constant infall of material.
Note this reference simulation is for a Schwarzschild black hole and no Blandford-Znajek-driving is possible
(see below for rotating black holes).
This axial area stays almost free of magnetic flux until the very final stages of the simulation.
At this point in time, the accumulation of magnetic flux also leads to an increasing strength of the outflows from the 
inner part of the disk.
These outflows are rooted in the accretion disk and outside the accretion funnel as they are seen in Figure~\ref{fig:velocity1}.

Concerning the disk dynamics, we observe a mixture of positive and negative velocities, overall pointing towards a 
turbulent nature of the accretion disk.
Strong accretion is only detected inside the marginally stable orbit.

\begin{figure}
\centering
    \includegraphics[width=0.98\columnwidth]{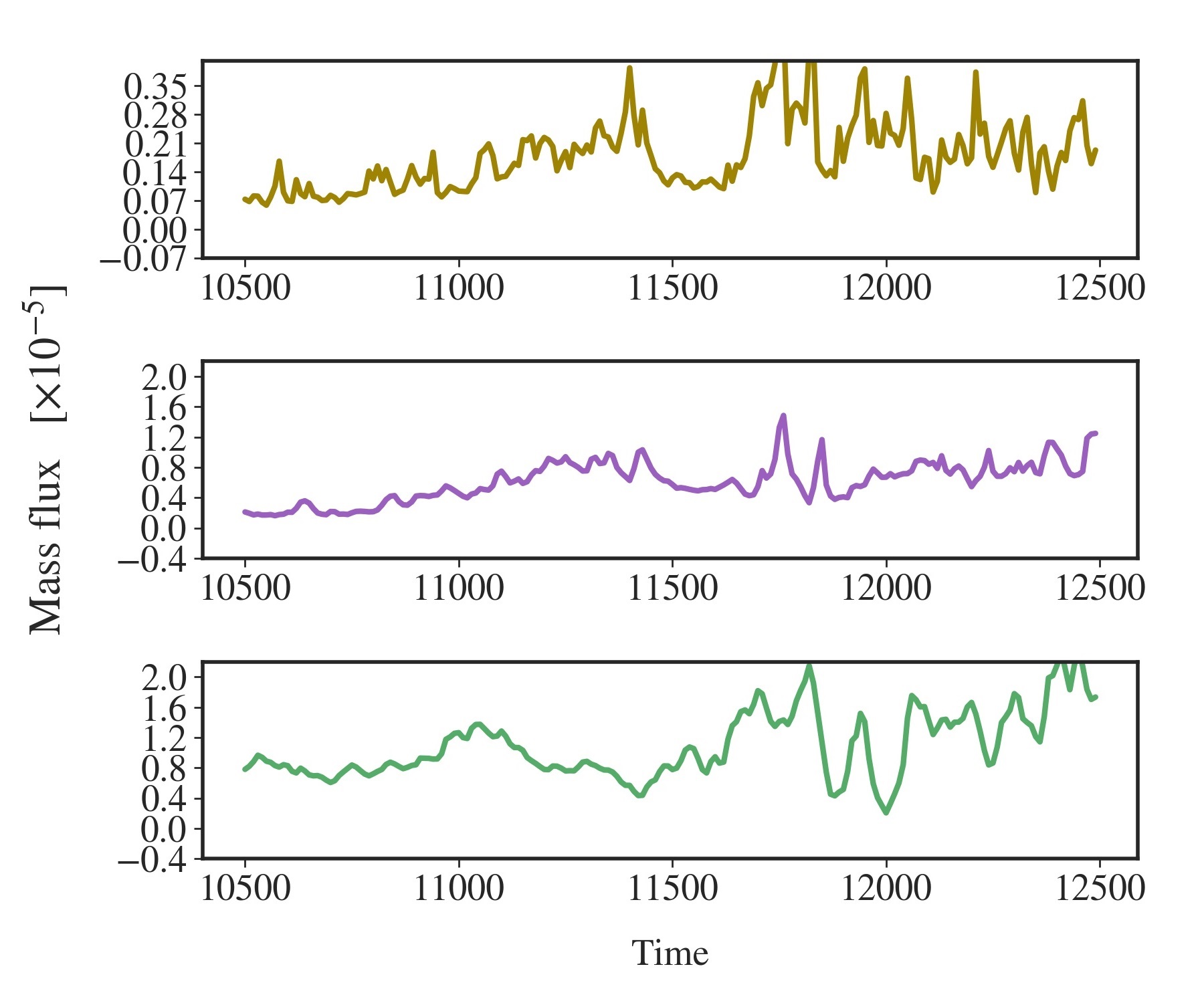}
    \includegraphics[width=0.98\columnwidth]{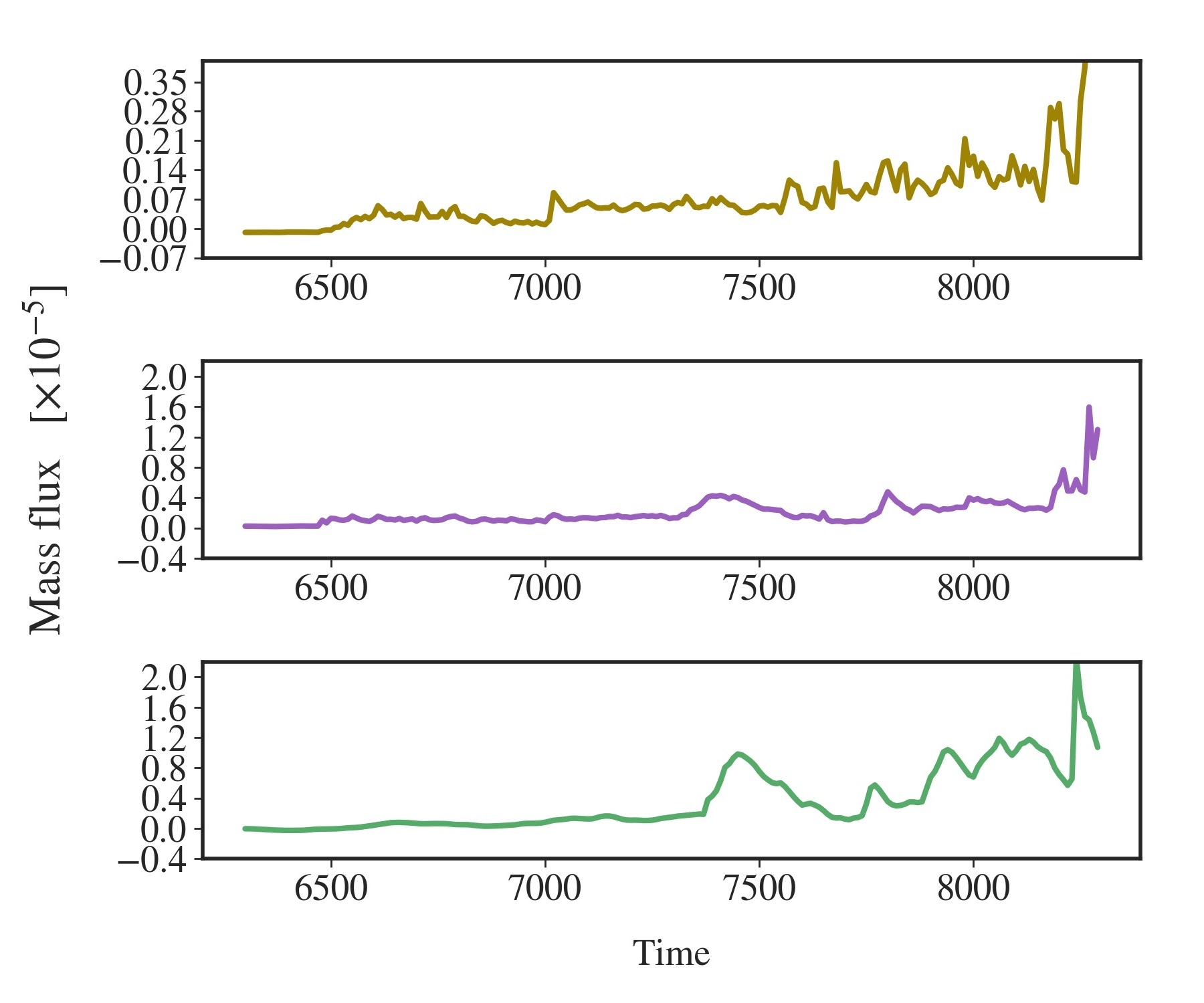}
\caption{Integrated normalized radial mass flux through surfaces of constant radius at $r \sim 32$ for simulations {\em sim0.1} 
(top) and {\em sim1.1} (bottom). 
The integration is done along three different circular sections, between
$0^{\circ} < \theta < 25^{\circ}$, $25^{\circ} < \theta < 65^{\circ}$, 
and $65^{\circ} < \theta < 80^{\circ}$ including the symmetric sections in the lower hemisphere.
Note the upper panels in the two groups of panels has tick marks almost an order of magnitude below the others.}
\label{fig:radialwind}
\end{figure}

Mow we want to quantify the fluxes carried by the outflows.
In Figure~\ref{fig:radialwind} we show the (normalized) radial mass flux integrated along the polar angle at 
radius $r\sim 32$ for simulations {\em sim0.1} and {\em sim1.1}.
These values express the rate at which material is flowing outwards (positive values) or flowing inwards 
(accretion, collapse) towards 
the black hole (negative values).

We may compare three different regions along the polar angle.
The first region is approximately $0^{\circ} < \theta < 25^{\circ}$, which is the funnel region close to the rotational axis. 
The second region is within $25^{\circ} < \theta < 65^{\circ}$ which represents the area where a high speed outflow may appear.
The third region is within $65^{\circ} < \theta < 80^{\circ}$, which is the area of a potential massive disk wind.
The respective areas of the lower hemisphere are also added to the integration of the mass fluxes.

The figure only shows the evolution of the mass flux for the late times of the simulation, this is the time when a disk wind is
established.
When comparing the numerical values, we see that simulation {\em sim0.1} shows stronger winds with an average mass 
flux in the second and third 
regions of $5.7\times 10^{-6}$ and $1.1\times 10^{-5}$.
In contrast, in simulation {\em sim1.1} we find mass fluxes that are only about 50\% of that, namely $2.5\times 10^{-6}$, and
$8.8\times10^{-6}$, respectively.

Note that the only difference in these simulations is the threshold for the quenching (see Table 1).
For {\em sim1.1} the dynamo is quenched only at a higher magnetization level, therefore producing a correspondingly stronger 
flux (compare also to Figure~\ref{fig:magneticflux_absB}).
We learn from this comparison that an accretion disk that carries a stronger magnetic flux, will eventually generate 
an outflow 
of lower mass flux. 
We emphasize, however, that simulation {\em sim1.1} terminates much earlier than simulation {\em sim0.1} due to strong 
magnetic field close to the black hole\footnote{This was possible due to the higher magnetization threshold we have 
chosen for {\em sim1.1}.}.
We suspect that if simulation {\em sim1.1} would not have crashed, a correspondingly stronger wind would have been formed.

We close this section by noting that wind is strong enough to affect the disk mass evolution, thus changing the slope of the disk 
mass evolution towards a larger mass loss, i.e. a more rapidly decreasing disk mass (see Figure~\ref{fig:accretionrate}, bottom).

\begin{figure}
\centering
    \includegraphics[width=0.98\columnwidth]{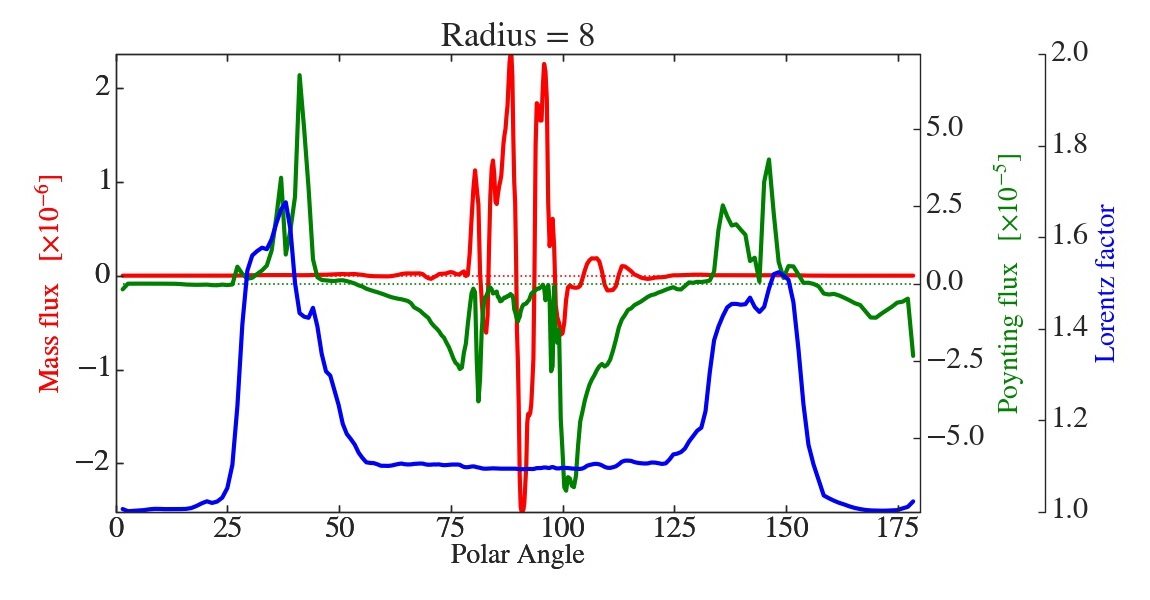} 
    \includegraphics[width=0.98\columnwidth]{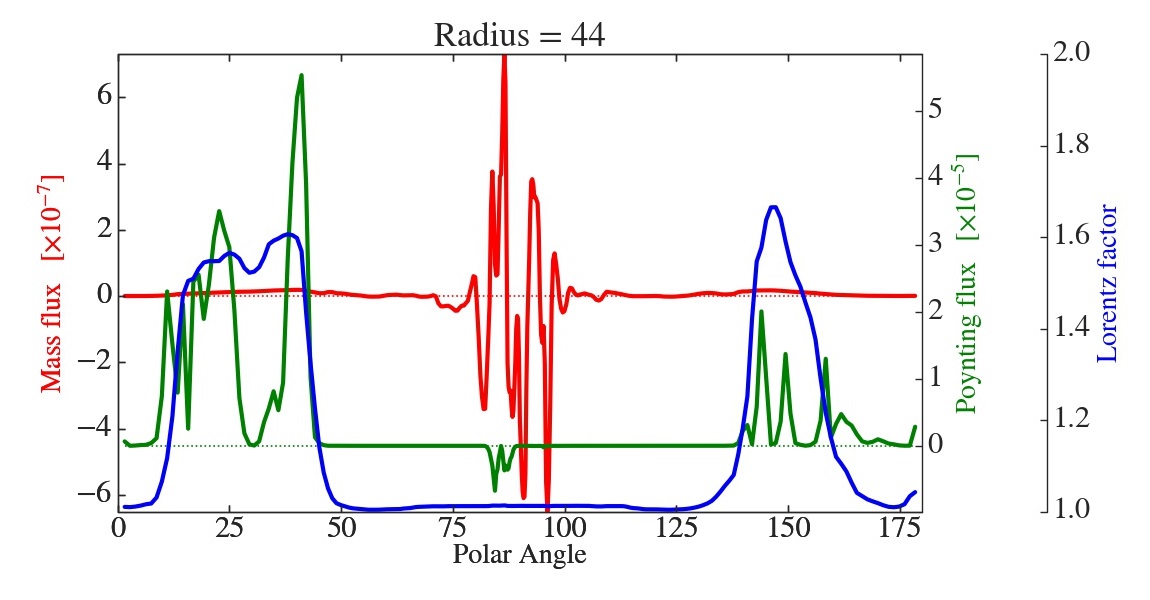}
\caption{Comparison of the angular distribution of normalized mass flux (red), Poynting flux per solid angle (green) 
    and Lorentz factor (blue) for simulation {\em sim0.1} at $t=7000$ at radius $r=8$ and $r=44$ for both hemispheres.
    Negative mass flux indicates accretion towards the black hole. 
    The Blandford-Znajek driven jet funnel is clearly distinguished by the peaks in Lorentz factor and electromagnetic energy flux.
    The mass flux increases with radius demonstrating a disk wind that is dominated by disk material.
    }
\label{fig:polarplots}
\end{figure}

%-----------------------------------------------------------------------------------------
\subsection{Generating a Blandford-Znajek jet}
\label{sec:BZjet}
Simulations with rotating black holes show the launching of strong outflows from the black hole magnetosphere,
especially in later stages when the magnetic field has engulfed the axial area.
This is a clear signature of Blandford-Znajek jets driven by the black hole ergosphere.
In Figure~\ref{fig:polarplots} we show the angular profile of mass flux, Poynting flux and Lorentz factor 
at radii $r=8,44$ and at $t=7000$ for simulation {\em sim0.1}.

\begin{figure*}
\centering
    \includegraphics[width=0.98\columnwidth]{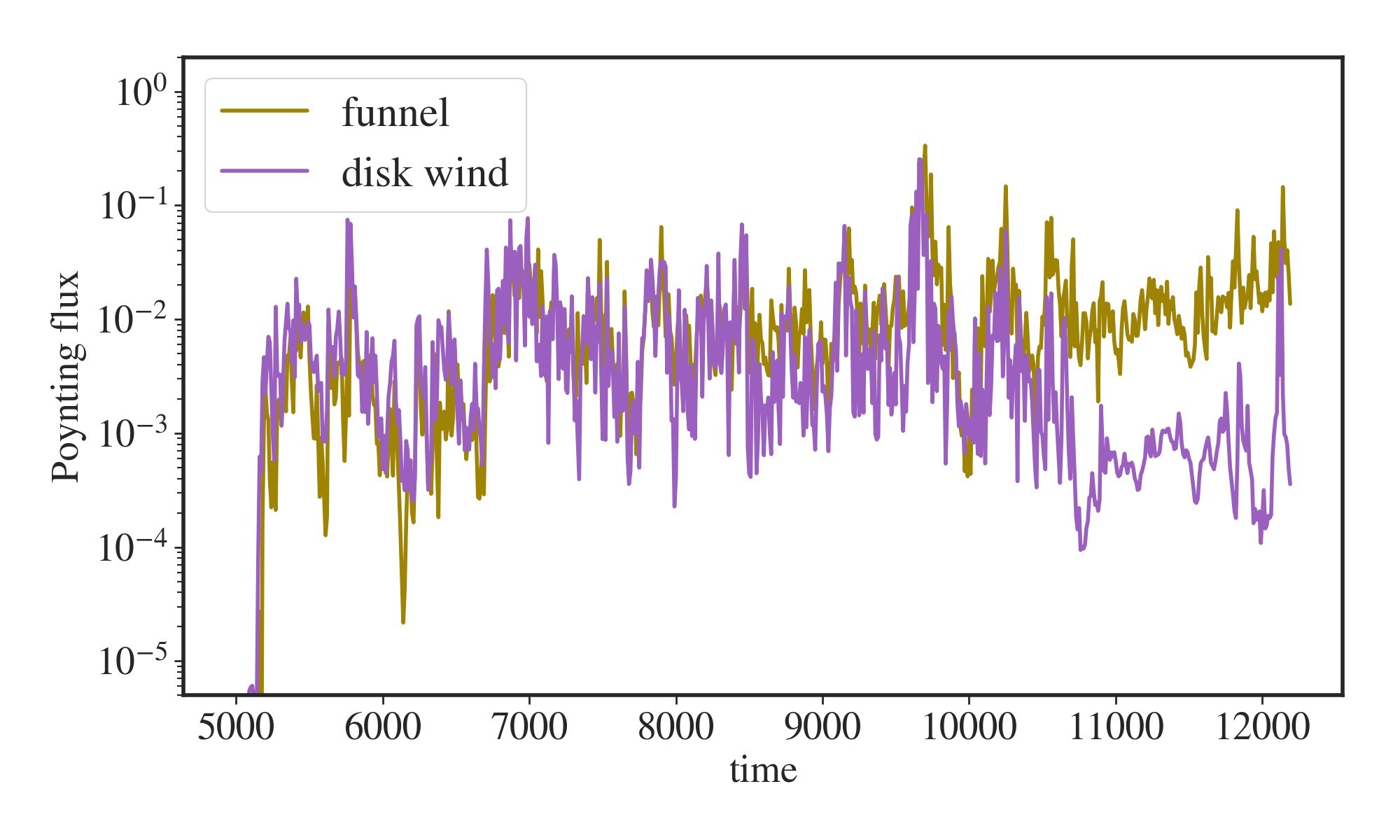} 
    \includegraphics[width=0.98\columnwidth]{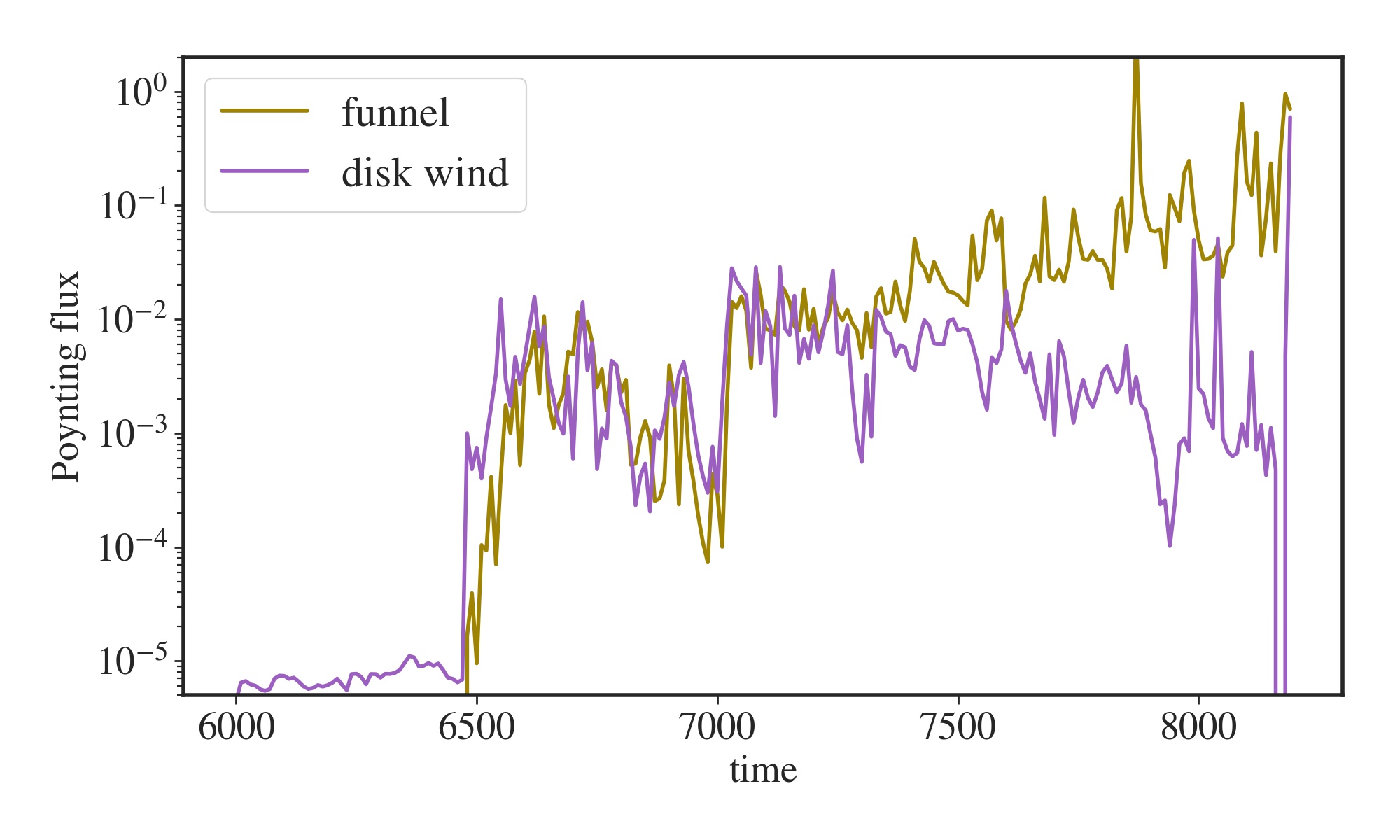}
    \includegraphics[width=0.98\columnwidth]{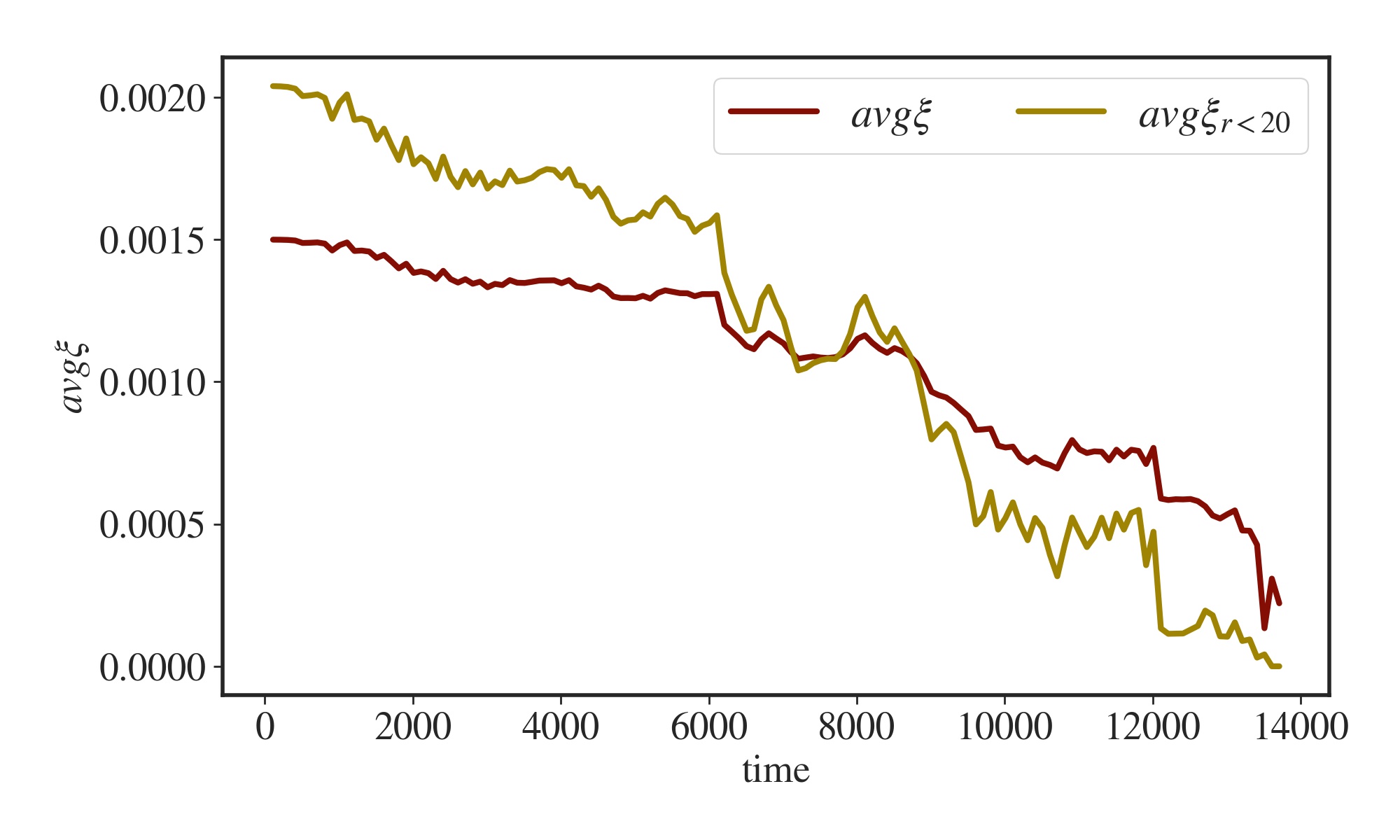}
    \includegraphics[width=0.98\columnwidth]{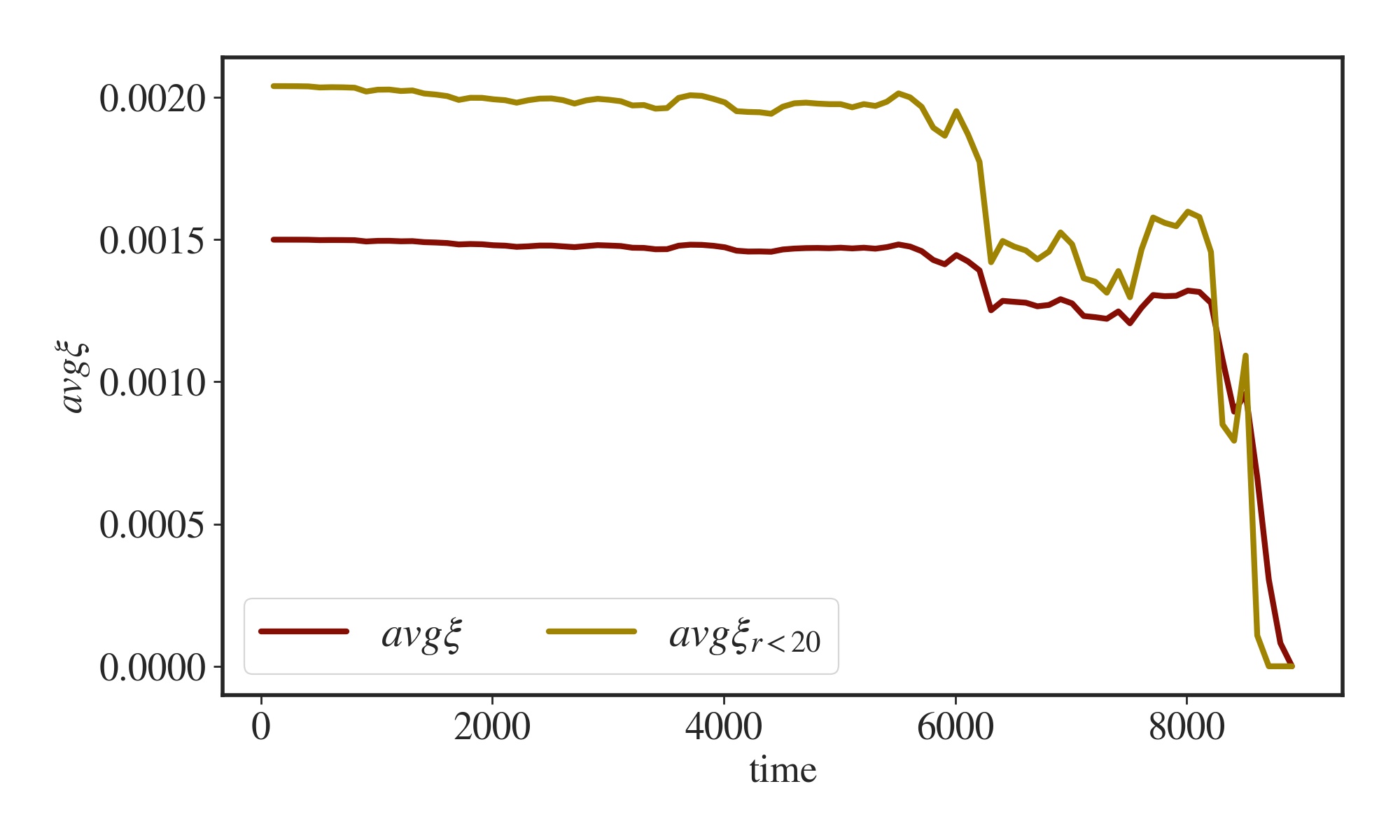}
\caption{Integrated Poynting flux {\it (top)} through surfaces of constant radius $r \sim 32$ for simulations 
{\em sim0.1} 
(left) and {\em sim1.1} (right), respectively. 
The integration is done along two segments, between 
$0^{\circ} < \theta < 25^{\circ}$ (funnel) and $25^{\circ} < \theta < 65^{\circ}$ (disk wind),
and their corresponding areas in the lower hemisphere, respectively.
The corresponding evolution of the absolute dynamo parameter $\xi$ {\it (bottom)},
averaged in space for the whole disk and the inner area $(r<20)$.
}
\label{fig:poyflux}
\end{figure*}

For the inner radius the mass flux has negative values around the equatorial plane 
which change into positive ones for slightly higher and lower angles.
Accretion towards the black hole is determined by the negative mass flux, while the mass outflow in the radial direction
is attributed to a low velocity wind from the inner disk and is expressed by the positive values.
For angles closer to the polar axis the mass flux decreases significantly. 
In this area the mass flux is determined by the floor density chosen for the simulation, a feature that is common to
all GRMHD simulations in the literature.
Note however, that in our case only the axial outflow is affected and not the disk wind that is loaded with disk material
and that carries a gas density is substantially higher than the floor values.
The Poynting flux remains largely negative for polar angles close to the equatorial plane.
It is positive for angles closer to the axis, indicative of a Poynting-dominated Blandford-Znajek jet.
Similarly, this is accompanied by increasing Lorentz factor towards the axis, again implying the launching
of a relativistic jet.

The picture becomes even more clear for the respective profiles derived for the larger radius. 
The Poynting flux is predominately positive in all directions with the higher values detected for the relativistic jet, 
similar for the Lorentz factor.
Note that this Poynting flux is essentially generated by the disk dynamo and either launched directly from the disk surface,
or, after advection of magnetic flux along the accretion disk, being further amplified by the black hole rotation.

The mass flux, however, gives a somewhat different picture.
It shows variations across the equatorial plane with quite large positive and negative deviations from an average value
(about more than 4 times lower than the variations in the inner disk).
At this point in time, at radius $r=44$ the disk has completed almost 4 orbits around the black hole.
This has to be compared to almost 45 disk orbits at $r=8$.
The much slower evolution of the outer parts of the disk, in combination with the entangled magnetic field results
in a much more turbulent accretion pattern of the outer disk parts.
Reconnection also plays a role as the entangled magnetic field may be anti-aligned (see also \citet{Vourellis2019} for a 
discussion of resistive effects for GRMHD jet launching). 
Note however, that the evolutionary time of the outflow is quite faster than the disk evolution, meaning that the 
observed outflow dynamics provide an instant tracer of the launching conditions at the foot point of the outflow.

In Figure~\ref{fig:poyflux} we show the evolution of integrated Poynting flux (integrated along circles of constant radius, $r=32$)
for simulations {\em sim0.1} and {\em sim1.1} hosted by a rotating black hole with $a=0.9$.
We concentrate on two regions.
One is the axial funnel area of $0^{\circ}<\theta<25^{\circ}$ where the relativistic jet launched by the black hole
magnetosphere is propagating.
The other is the disk wind area of $25^{\circ}<\theta<65^{\circ}$ where massive outflows from the disk are being 
launched.
The values of the symmetric areas in the lower hemisphere are taken into account as well.
We focus on the later stages of the simulations when the Poynting flux increases substantially.

The phase of strong Poynting flux begins approximately at the same time for both simulations.
However, the higher threshold in plasma-$\beta$ for the dynamo quenching allows simulation {\em sim0.1} to establish
a high Poynting flux for a much longer time.
In both simulations the Poynting flux measured in the jet funnel and in disk wind start with similar strength.
Magnetic flux is advected and when the area close the black hole and the axial region is more and more encompassed
by the dynamo-generated magnetic field, we see that the flux within the funnel increases while the flux of the disk wind decreases
(see the diverging lines in Fig.~\ref{fig:poyflux}, top panels).
Our interpretation is that magnetic flux that is first anchored in the disk, is advected to the funnel, thus increasing the funnel
flux and reducing the disk flux.
Especially for simulation {\em sim1.1} this reconfiguration process is more obvious. 
In Figure~\ref{fig:poyflux} (bottom) we show the evolution of the averaged absolute value of the
disk dynamo parameter $\xi$ for the same two simulations. 
For simulation {\em sim1.1}, for which the quenching of the dynamo starts around $\beta_{eq} = 100$, we notice the flat profile for the dynamo coincides with the absence of Poynting flux until the sudden increase of the Poynting flux.
Later, when the flux from the disk wind decreases $(t \simeq 7250)$ we find a slight increase in the
average dynamo parameter, especially for the one that includes the outer part of the disk.

These considerations are also supported by the time evolution of the magnetic flux of the disk and the black hole
that we have discussed above (see Figure~\ref{fig:magneticflux_absB}).
From the poloidal flux advected to the ergosphere, the black hole rotation will induce a toroidal magnetic field,
thus also increasing the Poynting flux.
Note that at this point in time the dynamo action has been reduced significantly due to the quenching 
(see Section~\ref{sec:quenching2}) and only little magnetic flux can be generated. 
Thus, the dynamo action is saturated and in balance with quenching.

%==================================================================================================
\section{Summary}
\label{sec:sum}
In this paper, we have extended our resistive GRMHD code \texttt{rHARM3D} \citep{QQ1, Vourellis2019} implementing a mean-
field dynamo in order to study the generation of a large-scale magnetic field and its effect on the launching
of outflows.
We have focused on a setup considering a thin accretion disk around a black holes running a number of simulation with 
various thresholds for the dynamo quenching and also applying different Kerr parameters for the black hole rotation.

Our mean-field dynamo simulations for thin accretion disk are initiated with a weak seed field in radial direction
that is confined to the disk.
The diffusivity profile follows a plateau profile within the disk and quickly drops across the disk surface towards
an ideal MHD corona.
Similarly, the profile for the mean-field dynamo parameter $\xi$ follows a sinusoidal profile in vertical direction from the 
the equatorial plane and a $1/\sqrt{r}$ profile along radius.

In the following we summarize our results. 

(1) Our implementation of the mean-field dynamo is based on the work of \citet{BdZ2013} and is an extension of the 
purely resistive version of our code \citep{Vourellis2019}.
The dynamo parameter $\xi$ (corresponding to the common dynamo-alpha in the literature) is inserted into Ohm's law as 
a source of the magnetic field, and via the Maxwell's equations into an equation for the evolution of the electric 
field.
We also apply a quenching mechanism to mitigate the exponential increase of the magnetic field.

(2) We first set up a torus initial setup for the simulations.
When comparing our results to those of \citet{Bugli2014, Tomei2020MNRAS.491.2346T}, we find in general
good agreement in the sense that the dynamo produces similar field structures within the inner torus.
In addition to the published literature we (i) continue the torus magnetic field evolution to
longer times scales, thereby being able to follow the advection of magnetic flux towards the black hole
rotational axis, and (ii) also provide the magnetic field lines of the field generated by the torus dynamo.

(3) The magnetic field evolution of the torus shows a saturation without applying a classical dynamo
quenching for the $\alpha$-effect in the code  as was reported by \citet{Tomei2020MNRAS.491.2346T}.
We hypothesize that in this particular case the saturation of the dynamo process was established by
{\em reconnection} of the dynamo-produced tangled magnetic field.
We propose this as another channel of dynamo quenching that works self-consistently in a turbulent
state of MHD, as long as physical resistivity is considered.

(4) Overall, the dynamo works as expected from non-relativistic simulations with the magnetic field lines emerging from 
the disk interior and expanding into the disk corona as the disk magnetic energy increases rapidly already at the beginning of 
the simulation.
Dynamo action slows down mainly due to the quenching function we apply.

The poloidal magnetic field dominates over the toroidal component especially in the outer parts of the disk where 
the difference can be an order of magnitude.
The outer parts of the disk remain weakly magnetized (at the time steps investigated),
but the disk magnetization still slowly increases till the final simulation times, with the toroidal field component
again dominating.

(5) We have also examined the evolution of the large-scale magnetic flux from the black hole-disk system for 
simulations applying different quenching levels for the dynamo as well as for rotating and non-rotating black holes,
as the flux is an essential quantity for jet launching.
The time evolution of the magnetic flux basically follows the evolution of magnetic energy and it is naturally
affected by the quenching threshold.
Even though the first stage is almost identical to all simulations, the ones with lower quenching threshold in plasma-$\beta$ increase their
magnetic field faster in the later stages to the point that they also terminate faster.
Black hole rotation appears to induce a delay in the increase of magnetic flux and also
leads to the higher levels of magnetization.

(6) The hydrodynamic evolution of the disk changes with the growth of the magnetic field.
Initially, the radial structure of the seed field allows for episodic accretion until the strong, dynamo generated 
field has evolved.
The existence of a dynamo-generated, strong vertical field component allows for the gradual appearance of disk winds,
which in turn lead to a smoothing of the otherwise quite peaky accretion rate.
At the final stages of the simulations, the interplay between dynamo activity and dynamo 
quenching, together with the diffusion of magnetic flux,
allows to develop a stronger disk wind together with a relativistic axial jet.

(7) We have identified small-scale disk outflows that are launched during the earlier stages of the simulations 
from the inner part of the accretion disk and which are interrelated with the presence of a strong magnetic 
field.
These small-scale outflows maybe considered as disk flares.
At the later stages of the simulation strong disk winds are ejected along the poloidal magnetic field, being 
mainly supported by the magnetic pressure of the toroidal field component.
We do not find indication for Blandford-Payne driven disk winds in our setup and for the time scales considered, 
as the disk winds is kinematically dominated (as also in the found in the simulations
by \citet{Vourellis2019}.

(8) In our simulations considering rotating black holes, we observe an additional outflow structure that is a highly relativistic jet, apparently driven by 
the Blandford-Znajek mechanism.
This jet carries a large electromagnetic energy flux when it leaves the ergosphere into the axial regions above 
the poles of the black hole (jet funnel).
By measuring the Poynting flux both in the jet funnel and in the disk wind, we find that the flux in the jet 
funnel increases over time in expense of the magnetic flux in the disk wind.
While the dynamo is quenched during later simulation times, and, thus, the disk magnetic flux becomes saturated,
advection of magnetic flux along the disk towards the black hole leads to the high levels of Poynting flux we observe in the funnel.

In summary, we have applied, for the first time, the mean-field dynamo theory in the GRMHD context of thin 
accretion disks around black holes.
We have investigated the evolution of the dynamo-generated magnetic field, the accretion disk and the outflows
that are launched with the help of the evolved magnetic field.
In particular we discuss the structure and evolution of the poloidal magnetic flux.

We find that the dynamo-generated magnetic field does not follow a clear dipolar structure, in difference to
non-relativistic studies in the literature.  
Future work is needed to understand the mean-field dynamo action in GRMHD in more detail, in particular simulations
running for longer time scales.
The lack of a well-aligned, large scale dipolar field does not allow to launch strong disk winds 
via magneto-centrifugal driving.
However, when the dynamo-generated magnetic field reaches a critical level, it is well capable to
launch magnetic pressure driven disk winds and also highly relativistic jets from the black hole ergosphere.

%==========================================================================================
\acknowledgements
The authors are grateful to Scott Noble for the possibility to use the original HARM3D code for further 
development (resistivity, dynamo).
Preparatory work by Qian Qian was extremely helpful for completing this work.
C.V. thanks the International Max Planck Research School for Astronomy and Cosmic Physics
at the University of Heidelberg (IMPRS-HD) for funding.
C.V. also wants to thank Hans-Walter Rix for financial support.
We acknowledge a detailed report with many helpful and interesting suggestions by an unknown referee. 

%===========================================================================================
\bibliography{main}{}
\bibliographystyle{aasjournal}
%---------------------------------------------------------------
\end{document}